\documentclass[10pt,twocolumn]{article}
\usepackage[margin=1in,top=0.75in,columnsep=0.3in]{geometry}
\usepackage{amsmath,amssymb}
\usepackage{graphicx}
\usepackage{booktabs}
\usepackage{tabularx}
\usepackage{multirow}
\usepackage{makecell}
\usepackage[colorlinks=true,linkcolor=blue,citecolor=blue,urlcolor=blue]{hyperref}
\usepackage{xcolor}
\usepackage[T1]{fontenc}
\usepackage{microtype}
\usepackage{url}
\usepackage{caption}
\usepackage{subcaption}
\usepackage{enumitem}
\usepackage{tikz}
\usetikzlibrary{arrows.meta,positioning,decorations.pathreplacing}
\usepackage{placeins}
\usepackage{balance}
\DeclareMathOperator*{\argmax}{arg\,max}
\title{%
\Large\textbf{Beyond Binary Priorities: Multi-Tier SLA Scheduling}\\[4pt]
\Large\textbf{for Large Language Model Serving}
}
\author{%
  Anders Vestrum \quad Arya Raeesi \quad Hanna Roed \\[4pt]
\small UC Berkeley, EECS Department}
\date{}
\begin{document}
% ===========================================================
\maketitle
% ---- Abstract ----
\begin{abstract}
Modern LLM serving deployments must simultaneously satisfy heterogeneous
service-level objectives (SLOs) across a diverse population of user tiers,
ranging from latency-critical API calls to background batch processing.
Llumnix~\cite{sun2024llumnix} introduced a dynamic, migration-capable
multi-instance scheduler for LLM inference that achieves load balancing,
defragmentation, prioritization, and auto-scaling through a unified
``freeness'' metric. However, Llumnix's priority model is restricted to two
levels (high and normal), an abstraction too coarse to express the richer SLA
classes common in production deployments. In this work, we extend Llumnix's
priority model to support an arbitrary number of tiers and evaluate the effects
of this extension under three realistic priority distributions (uniform,
Gaussian, enterprise) using Vidur~\cite{agrawal2024vidur}, a high-fidelity LLM
inference simulator. We implement per-tier headroom with exponential decay,
tier-aware dispatch ordering, and the full Llumnix migration pipeline inside
Vidur's hierarchical scheduling framework. We compare our extended scheduler
against INFaaS~\cite{romero2021infaas} (global routing baseline),
vLLM~\cite{kwon2023vllm}, Orca~\cite{yu2022orca}, and
Sarathi-Serve~\cite{agrawal2024sarathiserve} (per-replica baselines), sweeping
priority levels from~1 to~10. Our experiments demonstrate that four priority
tiers yields the best cost-effectiveness tradeoff, achieving prefill mean
speedups of up to $8.3{\times}$ and end-to-end P99 speedups of up to
$3.1{\times}$ over INFaaS with cost-per-latency improvements of 46--68\%,
while preserving strong SLO differentiation across tiers. We further show that
the system sustains these gains at 10 priority levels without tail latency
collapse, with overhead concentrated in the prefill phase.
\end{abstract}
% ===========================================================
\section{Introduction}
\label{sec:intro}
% ===========================================================
The rapid growth of large language model (LLM) deployments has brought
fundamentally new challenges to inference serving infrastructure. Unlike
classical deep learning inference, autoregressive LLM generation is highly
unpredictable: a single request's token count, memory footprint, and execution
time cannot be determined until generation
completes~\cite{sun2024llumnix}. This inherent heterogeneity, combined with
bursty arrival patterns and diverse latency requirements across users, renders
traditional single-instance, FCFS-based serving
insufficient~\cite{yu2022orca}.

State-of-the-art single-instance schedulers such as vLLM~\cite{kwon2023vllm},
Orca~\cite{yu2022orca}, Sarathi-Serve~\cite{agrawal2024sarathiserve}, and
FasterTransformer address per-replica throughput and memory efficiency, but
they are fundamentally designed for homogeneous workloads within a single
instance. They dispatch requests using simple round-robin or FCFS policies at
the cluster level, and once a request is placed on a replica, it stays there.
This ``one-shot'' dispatching is structurally incapable of reacting to memory
fragmentation, cross-instance load imbalances, or the need to re-isolate
high-priority requests that arrived after an instance became overloaded.

Llumnix~\cite{sun2024llumnix} addresses these limitations by introducing a
\emph{runtime rescheduler} that can live-migrate requests across instances with
near-zero downtime. It unifies load balancing, defragmentation, priority
isolation, and auto-scaling under a single ``freeness'' metric, and demonstrates
substantial latency and cost improvements over INFaaS~\cite{romero2021infaas}
as a baseline global scheduler. However, Llumnix's priority model remains
binary: requests are classified as either ``high priority'' or ``normal,'' with
a single fixed headroom value controlling isolation between the two classes.
This coarse granularity is insufficient for production LLM deployments, where
providers commonly differentiate across three to five SLA tiers (e.g., platinum,
gold, silver, standard, free-tier) with distinct latency targets.

This paper asks: \emph{what is the right number of priority tiers for a
migration-capable LLM serving system, and how should isolation headroom be
allocated across them?} We address this question by extending Llumnix to support
$K \geq 2$ priority levels, assigning each tier its own headroom budget via
exponential decay, and evaluating the system across three workload distributions
at scale.

\noindent Our primary contributions are:
\begin{enumerate}[leftmargin=*,topsep=2pt,itemsep=1pt]
\item \textbf{A multi-tier extension of Llumnix} integrated into the Vidur
    simulator, supporting up to 10 priority levels with per-tier headroom,
    priority-aware dispatch ordering, and full migration support.
\item \textbf{A comprehensive evaluation framework} spanning three realistic
    priority distributions (uniform, Gaussian, enterprise), two request volume
    scales (10K and 15K requests), and comparison against four baseline
    schedulers.
\item \textbf{An empirical characterization of the priority-complexity tradeoff},
    demonstrating that 4 priority tiers achieves the best cost-efficiency point,
    with diminishing isolation gains and increasing overhead beyond this threshold.
\item \textbf{Insights into the interaction between priority granularity and
    system load}, showing that the benefits of fine-grained prioritization are
    most pronounced at moderate loads where headroom can be honored, and diminish
    near saturation.
\end{enumerate}
% ===========================================================
\section{Background and Related Work}
\label{sec:background}
% ===========================================================
\subsection{LLM Inference Fundamentals}
\label{sec:bg:fundamentals}
LLM inference consists of two phases: \emph{prefill}, where the model processes
all input tokens in parallel to produce the initial KV-cache state, and
\emph{decode}, where tokens are generated autoregressively one at a time. The
prefill phase is compute-bound and bursty; the decode phase is
memory-bandwidth-bound and iterative. This two-phase structure creates a natural
tension: batching more prefill requests improves throughput but increases
time-to-first-token (TTFT), while prioritizing decode continuity reduces
time-between-tokens (TBT) but starves new
arrivals~\cite{agrawal2024sarathiserve}.

The KV-cache, which stores intermediate attention states for all active requests,
is the dominant memory consumer in LLM inference. As output length is unknown a
priori, KV-cache allocation grows dynamically throughout a request's lifetime.
Memory fragmentation, where logical contiguous allocations become physically
scattered, was a major throughput bottleneck until
PagedAttention~\cite{kwon2023vllm} introduced block-based virtual memory
management for KV caches, reducing waste to under 4\% in most cases.

\subsection{Per-Replica Scheduling}
\label{sec:bg:replica}
\noindent\textbf{Orca}~\cite{yu2022orca} introduced \emph{iteration-level
scheduling} (continuous batching), making batching decisions at each transformer
step rather than at request granularity. This eliminates head-of-line blocking
where long requests block short ones, and is now the standard scheduling
paradigm adopted by all modern serving systems.

\noindent\textbf{vLLM}~\cite{kwon2023vllm} pairs continuous batching with
PagedAttention to achieve near-zero KV-cache waste. It maintains a FCFS-ordered
waiting queue and a preemption policy (swap or recompute) for when memory
pressure forces request eviction.

\noindent\textbf{Sarathi-Serve}~\cite{agrawal2024sarathiserve} further decouples
the prefill-decode throughput-latency tradeoff by introducing \emph{chunked
prefills}, splitting long prefill sequences into equal-sized chunks interleaved
with ongoing decode steps, together with stall-free scheduling that prevents
decode stalls during prefill admission. This achieves up to $5.6{\times}$
improvement in serving capacity under pipeline parallelism.

\noindent\textbf{vAttention}~\cite{prabhu2024vattention} proposes an alternative
to PagedAttention that retains contiguous virtual address space for KV caches
while still achieving fragmentation-free physical allocation via CUDA virtual
memory APIs, improving serving throughput by up to $1.23{\times}$ over
block-based systems.

\subsection{Multi-Instance and Cluster-Level Scheduling}
\label{sec:bg:cluster}
\noindent\textbf{INFaaS}~\cite{romero2021infaas} is a model-less inference
serving system that automatically selects model variants and hardware per query.
It uses a cost-based routing function combining queue-depth estimation,
service-time prediction, and state-machine penalties (ACTIVE, OVERLOADED,
INTERFERED, INACTIVE) to route requests to the best available replica. While it
handles replica health and SLO tracking, it lacks live migration and cannot
react to per-request memory dynamics.

\noindent\textbf{AlpaServe}~\cite{li2023alpaserve} demonstrates that model
parallelism can serve as a multiplexing mechanism across heterogeneous model
requests, statistically sharing device capacity across bursty workloads. It
processes requests at up to $10{\times}$ higher rates while meeting P99 SLOs
for over 99\% of requests.

\noindent\textbf{MemServe}~\cite{hu2024memserve} introduces an elastic
distributed memory pool (MemPool) that manages KV caches across serving
instances, combining context caching with disaggregated inference and a
locality-aware scheduler for cache reuse maximization.

\noindent\textbf{Llumnix}~\cite{sun2024llumnix} combines live migration,
virtual usage tracking, and a unified freeness metric to unify load balancing,
defragmentation, priority isolation, and auto-scaling into a single scheduler.
It achieves up to $5.5{\times}$ prefill P99 speedup and 36\% cost reduction
over INFaaS. Our work directly extends this system.

\subsection{SLO-Aware and Priority-Aware Serving}
\label{sec:bg:slo}
\noindent\textbf{SCORPIO}~\cite{tang2025scorpio} introduces SLO heterogeneity
as a first-class scheduling concern, using TTFT Guard and TPOT Guard mechanisms
to maximize goodput under application-specific latency targets, improving
goodput by up to $14.4{\times}$ over prior baselines.

\noindent\textbf{SLOs-Serve}~\cite{chen2025slosserve} applies multi-SLO dynamic
programming to continuously optimize token allocations across stage-specific
constraints, combining chunked prefill with speculative decoding to achieve
$2.2{\times}$ per-GPU serving capacity gains.

\noindent\textbf{FlowPrefill}~\cite{hsieh2026flowprefill} tackles head-of-line
blocking in TTFT SLOs by decoupling preemption granularity from scheduling
frequency through operator-level preemption and event-driven scheduling,
improving maximum goodput by up to $5.6{\times}$ on production traces.

\noindent\textbf{SGDRC}~\cite{zhang2024sgdrc} addresses GPU-level priority
coexistence via software-defined dynamic allocation of VRAM bandwidth and
compute for latency-sensitive versus best-effort inference workloads, achieving
99\% average SLO attainment.

Prior work on multi-level feedback queues (MLFQ) and priority-based preemption
from classical scheduling theory~\cite{corbato1962mlfq} motivates our design
choice of exponential headroom decay, which mirrors the diminishing isolation
requirements at lower priority tiers while protecting system efficiency.

\subsection{Simulation for LLM Serving}
\label{sec:bg:simulation}
\noindent\textbf{Vidur}~\cite{agrawal2024vidur} is a large-scale, high-fidelity
LLM inference simulator from Microsoft Research that predicts latency,
throughput, model FLOPs utilization (MFU), and memory usage from profiled
operator runtimes. Its Vidur-Search component can identify optimal deployment
configurations for LLaMA2-70B in approximately 1 CPU-hour, compared to roughly
42{,}000 GPU-hours for real hardware search. Vidur includes baseline
implementations of vLLM, Orca, and Sarathi-Serve scheduling policies, providing
a fair, reproducible evaluation environment without requiring access to physical
GPU clusters.

% ===========================================================
\section{System Design}
\label{sec:design}
% ===========================================================
\subsection{Architecture Overview}
\label{sec:design:arch}
\begin{figure*}[t]
\centering
\begin{tikzpicture}[>=Stealth]

\tikzset{
  gsbox/.style={
    rectangle, rounded corners=6pt,
    draw=blue!65!black, fill=blue!9, line width=1.5pt,
    minimum width=4.2cm, minimum height=1.3cm,
    align=center, font=\small
  },
  llbox/.style={
    rectangle, rounded corners=5pt,
    draw=orange!75!black, fill=orange!9, line width=1.1pt,
    minimum width=2.6cm, minimum height=1.15cm,
    align=center, font=\small
  },
  prbox/.style={
    rectangle, rounded corners=3pt,
    draw=gray!45, line width=0.9pt,
    minimum width=2.2cm, minimum height=0.52cm,
    align=center, font=\footnotesize
  }
}

%% ===== PRIORITY INPUTS =====
\node[font=\fontsize{7.5}{9}\selectfont, gray!60!black]
  at (0, 1.35) {\textit{priority streams}};
\node[prbox, fill=red!17]    (p0) at (0,  0.75) {$P_0$: critical};
\node[prbox, fill=yellow!24] (p1) at (0,  0.12) {$P_1$: high};
\node[gray!60, font=\normalsize] at (0, -0.52) {$\vdots$};
\node[prbox, fill=gray!13]   (pk) at (0, -1.12) {$P_{K-1}$: background};

%% ===== GLOBAL SCHEDULER =====
\node[gsbox] (GS) at (5.0, -0.15)
  {\textbf{Llumnix}\\[3pt]\textbf{Global Scheduler}};

%% Priority streams → GS (converge to GS.west)
\draw[->, gray!50!black, thin] (p0.east) -- (GS.west);
\draw[->, gray!50!black, thin] (p1.east) -- (GS.west);
\draw[->, gray!50!black, thin] (pk.east) -- (GS.west);

%% ===== LLUMLETS =====
\node[llbox] (L1) at (1.6, -3.4)
  {\textbf{Llumlet 1}\\[2pt]{\scriptsize Replica 1}};
\node[llbox] (L2) at (5.0, -3.4)
  {\textbf{Llumlet 2}\\[2pt]{\scriptsize Replica 2}};
\node[gray!55, font=\normalsize] at (7.05, -3.4) {$\cdots$};
\node[llbox] (LN) at (9.2, -3.4)
  {\textbf{Llumlet $N$}\\[2pt]{\scriptsize Replica $N$}};

%% ===== DISPATCH  GS → Llumlets (blue solid) =====
%%
%% Each Llumlet gets its own departure point on GS.south
%% (spread -14pt, -4pt, +14pt) and its own out/in angles,
%% so the six arrows fan out cleanly without piling up.
%%
%% L1 — departs GS left-of-centre, curves down-left to L1
\draw[->, blue!70!black, line width=1.2pt]
  ([xshift=-14pt]GS.south) to[out=235, in=88] (L1.north);
%% L2 — straight down from centre, label on left side
\draw[->, blue!70!black, line width=1.2pt]
  ([xshift=-4pt]GS.south) -- ([xshift=-4pt]L2.north)
  node[pos=0.48, left=2pt, font=\fontsize{7}{8.5}\selectfont,
       blue!70!black] {dispatch};
%% LN — departs GS right-of-centre, curves down-right to LN
\draw[->, blue!70!black, line width=1.2pt]
  ([xshift=14pt]GS.south) to[out=305, in=92] (LN.north);

%% ===== FREENESS  Llumlets → GS (orange dashed) =====
%%
%% Use in/out angles offset ~15° from the dispatch arcs so the
%% two lanes run side-by-side rather than on top of each other.
%%
%% L1 — departs L1 at 65°, arrives at GS right-of-L1-zone
\draw[->, orange!80!black, dashed, line width=1.2pt]
  (L1.north) to[out=65, in=248] ([xshift=-6pt]GS.south);
%% L2 — straight up from centre, label on right side
\draw[->, orange!80!black, dashed, line width=1.2pt]
  ([xshift=4pt]L2.north) -- ([xshift=4pt]GS.south)
  node[pos=0.48, right=2pt, font=\fontsize{7}{8.5}\selectfont,
       orange!80!black] {freeness};
%% LN — departs LN at 115°, arrives at GS left-of-LN-zone
\draw[->, orange!80!black, dashed, line width=1.2pt]
  (LN.north) to[out=115, in=292] ([xshift=6pt]GS.south);

%% ===== MIGRATION (teal bidirectional dashed) =====
%%
%% L1 ↔ L2: clear bidirectional arrow with label
\draw[<->, teal!58!black, dashed, line width=1pt]
  (L1.east) -- (L2.west)
  node[midway, above, font=\fontsize{7}{8.5}\selectfont,
       teal!65!black] {migrate};
%% L2 ↔ dots: left stub, bidirectional
\draw[<->, teal!55!black, dashed, line width=0.9pt]
  (L2.east) -- (6.60, -3.4);
%% dots ↔ LN: right stub drawn left-to-right so arrowheads
%% are consistent: ← at stub start (toward dots), → into LN
\draw[<->, teal!55!black, dashed, line width=0.9pt]
  (7.50, -3.4) -- (LN.west);

%% ===== LEGEND =====
\begin{scope}[shift={(11.0, 0.25)},
              font=\fontsize{8}{10}\selectfont]
  \node[anchor=west, font=\fontsize{8}{10}\selectfont\bfseries]
    at (0, 0.72) {Legend};
  \draw[->, blue!70!black, line width=1.2pt]
    (0, 0.22) -- (1.0, 0.22) node[right] {\ dispatch};
  \draw[->, orange!80!black, dashed, line width=1.2pt]
    (0,-0.22) -- (1.0,-0.22) node[right] {\ freeness report};
  \draw[<->, teal!58!black, dashed, line width=1pt]
    (0,-0.66) -- (1.0,-0.66) node[right] {\ migration};
\end{scope}

\end{tikzpicture}
\caption{Two-level scheduling architecture. Priority-tagged request
  streams $P_0$ (critical) through $P_{K-1}$ (background) enter the
  Llumnix global scheduler, which dispatches them to $N$ Llumlet replica
  schedulers and receives freeness reports in return. Bidirectional
  migration links support live KV-cache transfer between replicas.}
\label{fig:architecture}
\end{figure*}
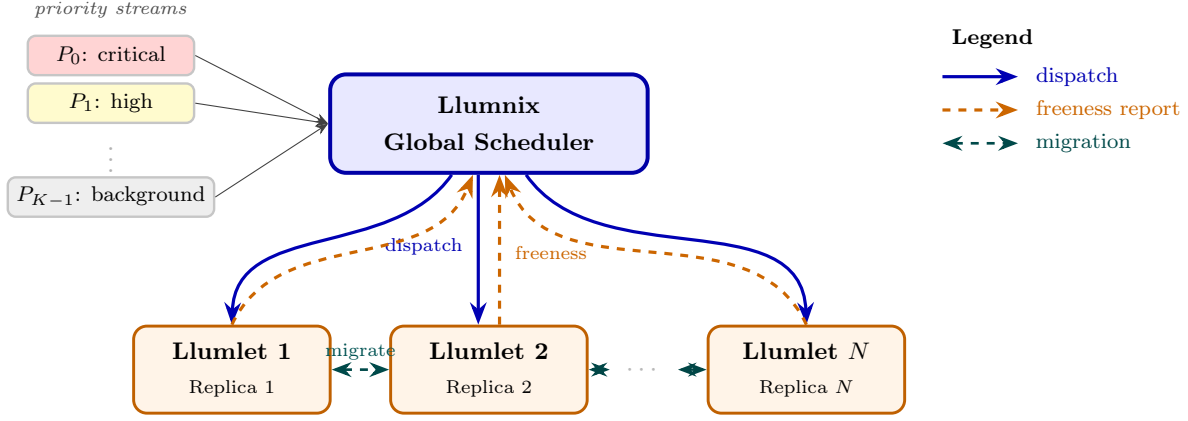

Our system follows Llumnix's two-level architecture
(Figure~\ref{fig:architecture}) adapted for the Vidur simulation framework.
A \textbf{global scheduler} (Llumnix-style) sits atop multiple \textbf{local
replica schedulers} (Llumlet-style), one per simulated GPU instance.

The global scheduler is responsible for dispatching incoming requests to the
freest non-draining replica, deciding when and which requests to migrate,
issuing auto-scaling recommendations based on cluster-wide freeness, and
marking replicas for graceful draining and evacuation.

Each Llumlet (local replica scheduler) is responsible for maintaining a
priority-ordered request queue, computing per-request virtual usage including
priority headroom, managing block-level KV-cache allocation, and executing
multi-stage live migration of running requests.

\subsection{The Freeness Metric}
\label{sec:design:freeness}
The central abstraction in our system is the \emph{freeness} of a replica:
\begin{equation}
  F = \frac{M - \sum_{r} V(r)}{B}
\label{eq:freeness}
\end{equation}
where $M$ is the total KV-cache block capacity of the replica,
$\sum_{r} V(r)$ is the total virtual usage summed across all requests $r$
currently associated with the replica (including queued requests), and $B$ is
the current batch size (number of actively running requests). Freeness $F > 0$
indicates available capacity; $F < 0$ indicates the replica is overloaded
relative to its virtual commitments, triggering migration.

The virtual usage $V(r)$ for a request $r$ is computed according to
Algorithm~1 of~\cite{sun2024llumnix} with our extension for $K$-level
priorities:
\begin{equation}
\begin{aligned}
V(r) =
\begin{cases}
\text{req.demand}
  & \text{if } r \text{ is HoL} \\[2pt]
0
  & \text{if } r \text{ queued} \\[2pt]
\text{physUsage}(r) + {}\\
\quad H(r_p, i)
  & \text{if } r \text{ running} \\[2pt]
\infty
  & \text{if drain}
\end{cases}
\end{aligned}
\end{equation}
where the headroom function $H(p, \text{instance})$ returns the per-priority
headroom allocation described in Section~\ref{sec:design:priority}.

\begin{figure}[htbp]
\centering
\resizebox{\columnwidth}{!}{%
\begin{tikzpicture}[font=\footnotesize]
  %-- Horizontal stacked bar: total width = 10 units (= M blocks) --
  \fill[green!40!white]  (0,0) rectangle (3.5,1.4);
  \fill[blue!30!white]   (3.5,0) rectangle (5.0,1.4);
  \fill[orange!45!white] (5.0,0) rectangle (7.0,1.4);
  \fill[gray!18]         (7.0,0) rectangle (10.0,1.4);
  \draw[thick,black!75]  (0,0) rectangle (10.0,1.4);
  \foreach \x in {3.5,5.0,7.0}
    \draw[black!35,thin] (\x,0)--(\x,1.4);
  % Inside labels
  \node[font=\small,green!50!black] at (1.75,0.70) {phys$(r)$};
  \node[font=\small,blue!65!black,align=center] at (4.25,0.70) {HoL\\demand};
  \node[font=\small,orange!75!black] at (6.0,0.70) {$H_p$};
  \node[font=\small,gray!60!black] at (8.5,0.70) {free};
  % Percentage labels above bar
  \node[above,font=\scriptsize] at (1.75,1.4) {35\%};
  \node[above,font=\scriptsize] at (4.25,1.4) {15\%};
  \node[above,font=\scriptsize] at (6.0,1.4) {20\%};
  \node[above,font=\scriptsize] at (8.5,1.4) {30\%};
  % Virtual usage brace (sits just below the bar)
  \draw[decorate,decoration={brace,amplitude=5pt,mirror},
        thick,black!70] (0,-0.18)--(7.0,-0.18)
    node[midway,below=6pt,font=\scriptsize] {$\textstyle\sum_r V(r)$};
  % Total M arrow (lower, clear of brace label)
  \draw[<->,thin,gray!55] (0,-0.90)--(10.0,-0.90);
  \node[below,font=\scriptsize] at (5.0,-0.90) {$M$ total KV capacity};
  % Legend row 1
  \node[font=\scriptsize,anchor=west] at (0,-1.85)
    {\textcolor{green!55!black}{$\blacksquare$}~phys$(r)$: running KV footprint \quad
     \textcolor{blue!65!black}{$\blacksquare$}~HoL: head-of-line demand};
  % Legend row 2
  \node[font=\scriptsize,anchor=west] at (0,-2.55)
    {\textcolor{orange!75!black}{$\blacksquare$}~$H_p$: per-tier headroom (drives migration) \quad
     \textcolor{gray!60!black}{$\blacksquare$}~free capacity};
  % Freeness formula
  \node[font=\small,align=center] at (5.0,-3.45)
    {$F = \bigl(M - \textstyle\sum_r V(r)\bigr)\,/\,B$};
\end{tikzpicture}%
}
\caption{Components of virtual usage in the freeness metric. The stacked bar
  shows a single replica's KV memory partitioned into running-request
  allocations, head-of-line demand, priority headroom, and free space.
  Freeness $F$ is positive when virtual usage leaves slack, and goes
  negative when headroom plus demand exceed $M$, triggering migration.}
\label{fig:virtual_usage}
\end{figure}

\subsection{Multi-Tier Priority Model}
\label{sec:design:priority}
The original Llumnix paper defines headroom as a single scalar $h$ such that
the headroom contribution for a high-priority request is
$h / \text{numHighPriorityRequests}$ and zero for normal requests. This
ensures that the freeness of a heavily loaded replica appears artificially
lower to the global scheduler, discouraging further normal-priority dispatch
while high-priority requests remain.

We generalize this to $K$ priority tiers. Each tier
$p \in \{0, 1, \ldots, K-1\}$ receives a dedicated headroom budget $H_p$
computed via exponential decay:
\begin{equation}
  H_p = M \cdot h_{\max} \cdot e^{-\lambda p}
\label{eq:headroom}
\end{equation}
where $M$ is the replica's block capacity, $h_{\max} = 0.20$ (20\% of
capacity reserved for tier~0), and $\lambda$ is chosen such that
$H_{K-1} \approx 0$. The total virtual headroom contributed to a replica's
freeness calculation is:
\begin{equation}
  V_{\text{headroom}} = \sum_{p=0}^{K-1} H_p \cdot
    \mathbf{1}\!\left[\text{numRequests}(p) > 0\right]
\label{eq:total_headroom}
\end{equation}
Note that the full budget $H_p$ is charged whenever \emph{any} request of
priority~$p$ is present, not per request, ensuring that a single critical
request can effectively block normal-priority dispatch to an overloaded
replica.

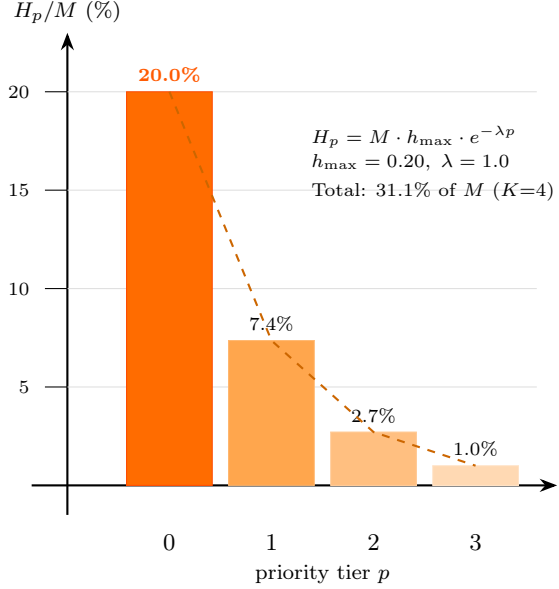
\begin{figure}[htbp]
\centering
% xscale/yscale give independent axis scaling without distorting bar shapes.
% xscale=1.35: 5 x-units -> ~6.75 cm (fits column after y-axis margin)
% yscale=0.26:  20 y-units -> ~5.2 cm tall  => natural ~7x5 cm figure
\begin{tikzpicture}[xscale=1.35, yscale=0.26, font=\footnotesize, >=Stealth]
  \def\hw{0.42}

  % Axes – stop arrow at x=4.8 so label stays within column
  \draw[->,thick] (-0.35,0) -- (4.8,0);
  % x-axis label centred below tick labels (clear of the arrow tip)
  \node[font=\footnotesize] at (2.5,-4.5) {priority tier $p$};
  \draw[->,thick] (0,-1.5) -- (0,23.0)
    node[above,font=\footnotesize] {$H_p/M$ (\%)};

  % Y grid + ticks
  \foreach \y/\lab in {5/5,10/10,15/15,20/20} {
    \draw[gray!25,thin] (0,\y) -- (4.6,\y);
    \draw[thin] (-0.22,\y)--(0,\y);
    \node[left,font=\scriptsize] at (-0.28,\y) {\lab};
  }

  % p=0: 20.00%
  \fill[orange!85!red]      (1-\hw,0) rectangle (1+\hw,20.00);
  \draw[orange!60!red,thin] (1-\hw,0) rectangle (1+\hw,20.00);
  \node[above,font=\scriptsize,orange!70!red] at (1,20.00) {\textbf{20.0\%}};

  % p=1: 7.36%
  \fill[orange!70]      (2-\hw,0) rectangle (2+\hw,7.36);
  \draw[orange!55,thin] (2-\hw,0) rectangle (2+\hw,7.36);
  \node[above,font=\scriptsize] at (2,7.36) {7.4\%};

  % p=2: 2.71%
  \fill[orange!50]      (3-\hw,0) rectangle (3+\hw,2.71);
  \draw[orange!40,thin] (3-\hw,0) rectangle (3+\hw,2.71);
  \node[above,font=\scriptsize] at (3,2.71) {2.7\%};

  % p=3: 1.00%
  \fill[orange!30]      (4-\hw,0) rectangle (4+\hw,1.00);
  \draw[orange!22,thin] (4-\hw,0) rectangle (4+\hw,1.00);
  \node[above,font=\scriptsize] at (4,1.00) {1.0\%};

  % X tick labels
  \foreach \x/\lab in {1/0, 2/1, 3/2, 4/3}
    \node[below,font=\small] at (\x,-2.0) {\lab};

  % Dashed decay curve
  \draw[dashed,orange!80!black,thick]
    (1,20.00) -- (2,7.36) -- (3,2.71) -- (4,1.00);

  % Annotation: placed high and to the right, well above the 7.4% bar
  % Physical y = 14*0.26 = 3.64 cm  (tier-1 bar top = 7.36*0.26 = 1.91 cm)
  \node[font=\scriptsize,align=left,anchor=south west] at (2.3,14.0)
    {$H_p = M \cdot h_{\max} \cdot e^{-\lambda p}$\\[1pt]
     $h_{\max}=0.20,\;\lambda=1.0$\\[3pt]
     Total: $31.1\%$ of $M$ ($K{=}4$)};
\end{tikzpicture}
\caption{Per-tier headroom allocation under exponential decay ($K=4$,
  $h_{\max}=0.20$, $\lambda=1.0$). Tier~0 reserves 20\% of KV capacity;
  each subsequent tier decays by $e^{-\lambda}{\approx}37\%$.
  Total headroom across all tiers is 31.1\% of $M$.}
\label{fig:headroom}
\end{figure}

\subsection{Priority-Aware Dispatch}
\label{sec:design:dispatch}
The global scheduler maintains a sorted pending queue and dispatches requests
in strict priority order (tier~0 first), with FCFS ordering within each tier.
For each request, the dispatcher selects the replica with the highest freeness
($\argmax F$) that is not currently draining. If all replicas are draining (a
transient state during aggressive scale-in), the dispatcher falls back to the
least-loaded draining replica.

Two separate freeness values are maintained per replica. \textbf{Full freeness}
$F_{\text{full}}$ includes priority headroom and is used for dispatch and
migration targeting. \textbf{Normal-priority freeness} $F_{\text{normal}}$
excludes headroom and is used exclusively for auto-scaling decisions. The
distinction prevents artificial inflation of virtual usage from triggering
spurious scale-out events, matching the auto-scaling component described
in~\cite{sun2024llumnix}.

\subsection{Migration and Load Rebalancing}
\label{sec:design:migration}
The global scheduler periodically evaluates whether migration is warranted
using two criteria: a time-based interval (50\,ms in most experiments) and a
load imbalance threshold $\Delta F = \max(F) - \min(F) \geq \theta$ (default
$\theta = 0.3$).

When migration is triggered, the scheduler identifies \emph{source} replicas
(low freeness) and \emph{destination} replicas (high freeness), and instructs
each source Llumlet to select and migrate one request. Candidate selection
prioritizes queued requests (no KV state, cheap to migrate) and secondarily
targets low-priority running requests with small KV footprints. This minimizes
migration cost while maximally relieving pressure on overloaded replicas.

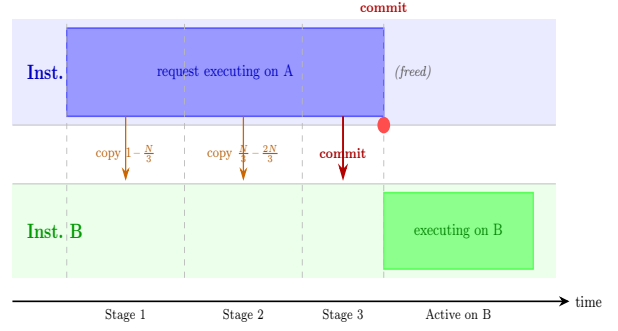
\begin{figure}[htbp]
\centering
% Taller resizebox to give the wider inter-lane gap room to breathe
\resizebox{\columnwidth}{4.4cm}{%
\begin{tikzpicture}[font=\footnotesize, >=Stealth]
  % Layout
  %   A lane : y = 3.0 .. 4.8   (1.8 units tall)
  %   gap    : y = 2.0 .. 3.0   (1.0 unit  – plenty of room for labels)
  %   B lane : y = 0.4 .. 2.0   (1.6 units tall)

  % --- Lane backgrounds ---
  \fill[blue!8]  (0,3.0) rectangle (12.0,4.8);
  \fill[green!8] (0,0.4) rectangle (12.0,2.0);
  \draw[gray!45,thin] (0,3.0) -- (12.0,3.0);
  \draw[gray!45,thin] (0,2.0) -- (12.0,2.0);

  % --- Lane labels ---
  \node[anchor=west,font=\normalsize\bfseries,blue!75!black]  at (0.2,3.9) {Inst.~A};
  \node[anchor=west,font=\normalsize\bfseries,green!60!black] at (0.2,1.2) {Inst.~B};

  % --- Execution bar on A (x: 1.2 .. 8.2) ---
  \fill[blue!40!white] (1.2,3.15) rectangle (8.2,4.65);
  \draw[blue!65,thick]  (1.2,3.15) rectangle (8.2,4.65);
  \node[font=\footnotesize,blue!75!black] at (4.7,3.9) {request executing on A};

  % --- Stage vertical dividers ---
  \foreach \t in {1.2,3.8,6.4,8.2}
    \draw[dashed,gray!55] (\t,0.4) -- (\t,4.8);

  % --- Copy arrows: from bottom of exec bar (3.15) to top of gap (3.0),
  %     continuing down to just above B lane top (2.1) ---
  \draw[->,orange!80!black,thick]   (2.5,3.15) -- (2.5,2.05);
  \draw[->,orange!80!black,thick]   (5.1,3.15) -- (5.1,2.05);
  \draw[->,red!70!black,very thick] (7.3,3.15) -- (7.3,2.05);

  % --- Inter-lane labels: centred in the gap (y = 2.52) ---
  \node[font=\scriptsize,align=center,orange!70!black] at (2.5,2.52)
    {copy $1$\,--\,$\tfrac{N}{3}$};
  \node[font=\scriptsize,align=center,orange!70!black] at (5.1,2.52)
    {copy $\tfrac{N}{3}$\,--\,$\tfrac{2N}{3}$};
  \node[font=\scriptsize,align=center,red!65!black]    at (7.3,2.52)
    {\textbf{commit}};

  % --- Active on B ---
  \fill[green!45!white] (8.2,0.55) rectangle (11.5,1.85);
  \draw[green!65,thick]  (8.2,0.55) rectangle (11.5,1.85);
  \node[font=\footnotesize,green!50!black] at (9.85,1.2) {executing on B};

  % --- Commit marker ---
  \fill[red!70] (8.2,3.0) circle (4pt);
  \node[above,font=\scriptsize,red!70!black] at (8.2,4.8) {\textbf{commit}};

  % --- (freed) label ---
  \node[font=\scriptsize,gray!55!black,anchor=west] at (8.3,3.9)
    {\textit{(freed)}};

  % --- Time axis ---
  \draw[->,thick] (0,0.0) -- (12.3,0.0)
    node[right,font=\footnotesize] {time};

  % Stage labels below axis
  \node[below,font=\scriptsize] at (2.5,0.0)  {Stage~1};
  \node[below,font=\scriptsize] at (5.1,0.0)  {Stage~2};
  \node[below,font=\scriptsize] at (7.3,0.0)  {Stage~3};
  \node[below,font=\scriptsize] at (9.85,0.0) {Active on B};

\end{tikzpicture}%
}
\caption{Multi-stage live migration of a running request's KV-cache from
  Instance~A to Instance~B. In each stage the Llumlet copies a fixed block
  chunk while the request continues executing on~A. After the final
  stage the request is committed to~B. Only a brief re-enqueue
  latency is incurred at the handoff; decode-phase tokens are generated
  throughout.}
\label{fig:migration}
\end{figure}

For \emph{running} requests, migration proceeds in multiple stages
(Figure~\ref{fig:migration}). Each stage copies a fixed number of KV blocks
from source to destination while the request continues to execute on the
source. After $\lceil B_r / B_{\text{stage}} \rceil$ stages, the final stage
commits the request to the destination. The request continues to generate
tokens throughout migration; only a brief re-enqueue latency is incurred at
the final handoff.

When a replica is draining (used for scale-in and graceful instance
shutdown), the Llumlet inserts an infinite-virtual-usage sentinel into the
freeness calculation, making the instance appear maximally overloaded. This
triggers continuous migration of all requests off the draining instance
without any new dispatch.

\subsection{Auto-Scaling}
\label{sec:design:autoscale}
Auto-scaling decisions use the cluster-average normal-priority freeness
$\bar{F}_{\text{normal}} = \frac{1}{N} \sum_i F_{\text{normal},i}$.
Scale-out is triggered when $\bar{F}_{\text{normal}} < -0.5$ (cluster
overloaded), and scale-in when $\bar{F}_{\text{normal}} > 1.5$ (cluster
underutilized). Using $F_{\text{normal}}$ rather than full freeness prevents
priority isolation from being misinterpreted as capacity shortage, which would
otherwise cause unnecessary scale-out as priority levels increase.

\subsection{INFaaS Baseline Implementation}
\label{sec:design:infaas}
We implement an INFaaS-style global scheduler~\cite{romero2021infaas} as the
primary comparison baseline. Our implementation captures INFaaS's core routing
mechanism via a cost-based dispatch function:
\begin{equation}
  \text{cost}(r) = \alpha \cdot \hat{q}_r + \beta \cdot \hat{s}_r +
\gamma \cdot \text{penalty}(r)
\label{eq:infaas_cost}
\end{equation}
where $\hat{q}_r$ is the estimated queue depth on replica~$r$, $\hat{s}_r$
is the expected service time estimated via EWMA, and $\text{penalty}(r)$ is a
state-based penalty (0 for ACTIVE, elevated for OVERLOADED and INTERFERED).
Unlike Llumnix, INFaaS performs no live migration and maintains no KV-aware
freeness; it relies entirely on load-balancing at dispatch time.

\FloatBarrier
% ===========================================================
\section{Implementation}
\label{sec:impl}
% ===========================================================
\subsection{Simulator Integration}
\label{sec:impl:simulator}
We implemented our extended Llumnix scheduler as a new scheduling policy
within the Vidur simulation framework~\cite{agrawal2024vidur}, forking both
the Vidur and Llumnix codebases.\footnote{Source code available at
\url{https://github.com/andersvestrum/llumnix\_sim}.} Vidur provides a discrete-event simulation
engine that models transformer operator latencies from real profiling data,
enabling accurate prediction of TTFT, TBT, and end-to-end request latency
without requiring GPU hardware access.

Our integration consists of three primary components.

\noindent\textbf{LlumletReplicaScheduler}
({\footnotesize\path{vidur/scheduler/replica_scheduler/llumlet_replica_scheduler.py}}):
A new replica scheduler implementing the full Llumlet logic, including
priority-ordered queueing, multi-component virtual usage, per-tier headroom
computation, block-level KV-cache management, and multi-stage migration state
machines.

\noindent\textbf{LlumnixGlobalScheduler}
({\footnotesize\path{vidur/scheduler/global_scheduler/llumnix_global_scheduler.py}}):
A new global scheduler implementing priority-aware dispatch (tier~0 first,
FCFS within tier), migration orchestration, auto-scaling signaling, and
draining. It communicates with Llumlets via \texttt{report\_freeness()} and
\texttt{begin\_migration\_to()} interfaces.

\noindent\textbf{InfaasGlobalScheduler}
({\footnotesize\path{vidur/scheduler/global_scheduler/infaas_global_scheduler.py}}):
The INFaaS-style baseline routing implementation with replica state tracking,
EWMA latency estimation, and cost-based request assignment.

\subsection{Priority Distribution Sampling}
\label{sec:impl:distributions}
We implement a \texttt{PrioritySampler}
({\footnotesize\path{vidur/utils/priority_sampler.py}}) supporting multiple request priority
distribution types. For our main experiments we use three distributions. The
\textbf{uniform} distribution assigns equal probability $1/K$ per tier,
stress-testing the scheduler with equal numbers of requests at each priority
level. The \textbf{Gaussian} distribution centers weights at tier
$\lfloor K/2 \rfloor$ with standard deviation $K/4$, reflecting a
middle-heavy workload. The \textbf{enterprise} distribution assigns
approximately 10\% of traffic to tier~0 (critical), $\sim$70\% to
mid-priority tiers, and 5--20\% to background tier $K{-}1$, modeling a
realistic production deployment where most traffic is standard quality.

Priority~0 is the highest priority (critical/interactive) and priority $K-1$
is the lowest (background/batch), consistent with Llumnix's original
convention.

\subsection{Request Length Distribution}
\label{sec:impl:lengths}
All experiments use a synthetic length distribution calibrated to represent a
realistic mix of LLM API traffic (Table~\ref{tab:lengths}). The distribution
is right-skewed to reflect the empirical observation that the majority of real
LLM API traffic consists of short conversational turns~\cite{sun2024llumnix}.

\begin{table}[h]
\centering
\caption{Request length distribution used in all experiments.}
\label{tab:lengths}
\footnotesize
\setlength{\tabcolsep}{5pt}
\begin{tabularx}{\columnwidth}{@{} c c X @{}}
\toprule
\textbf{Tokens} & \textbf{Prob.} & \textbf{Description} \\
\midrule
64--128  & $\sim$65\% & Short (chat, classification) \\
128--256 & $\sim$22\% & Medium (summarization) \\
256--384 & $\sim$10\% & Long (document analysis) \\
384--512 & $\sim$2\%  & Very long (code generation) \\
\bottomrule
\end{tabularx}
\end{table}

\subsection{Experiment Configuration}
\label{sec:impl:config}
All simulations run on 4 replica instances with aggregate query arrival rate
QPS~$= 1{,}250$ (Poisson-distributed) for the main priority distribution
sweeps. For the comparative scheduler benchmark
(Section~\ref{sec:eval:schedulers}), we use QPS~$= 10$ with 1{,}000 requests
to allow cleaner cross-scheduler comparison. All experiments were run using
120 parallelized CPU threads on Berkeley Research Computing infrastructure,
sweeping across 120 simulation configurations per distribution type.

\FloatBarrier
% ===========================================================
\section{Evaluation}
\label{sec:eval}
% ===========================================================
\subsection{Priority Differentiation Under Varying Tier Counts}
\label{sec:eval:differentiation}
We first evaluate whether our extended priority model successfully achieves
latency differentiation between tiers.

\begin{figure*}[t]
\centering
\includegraphics[width=\textwidth]{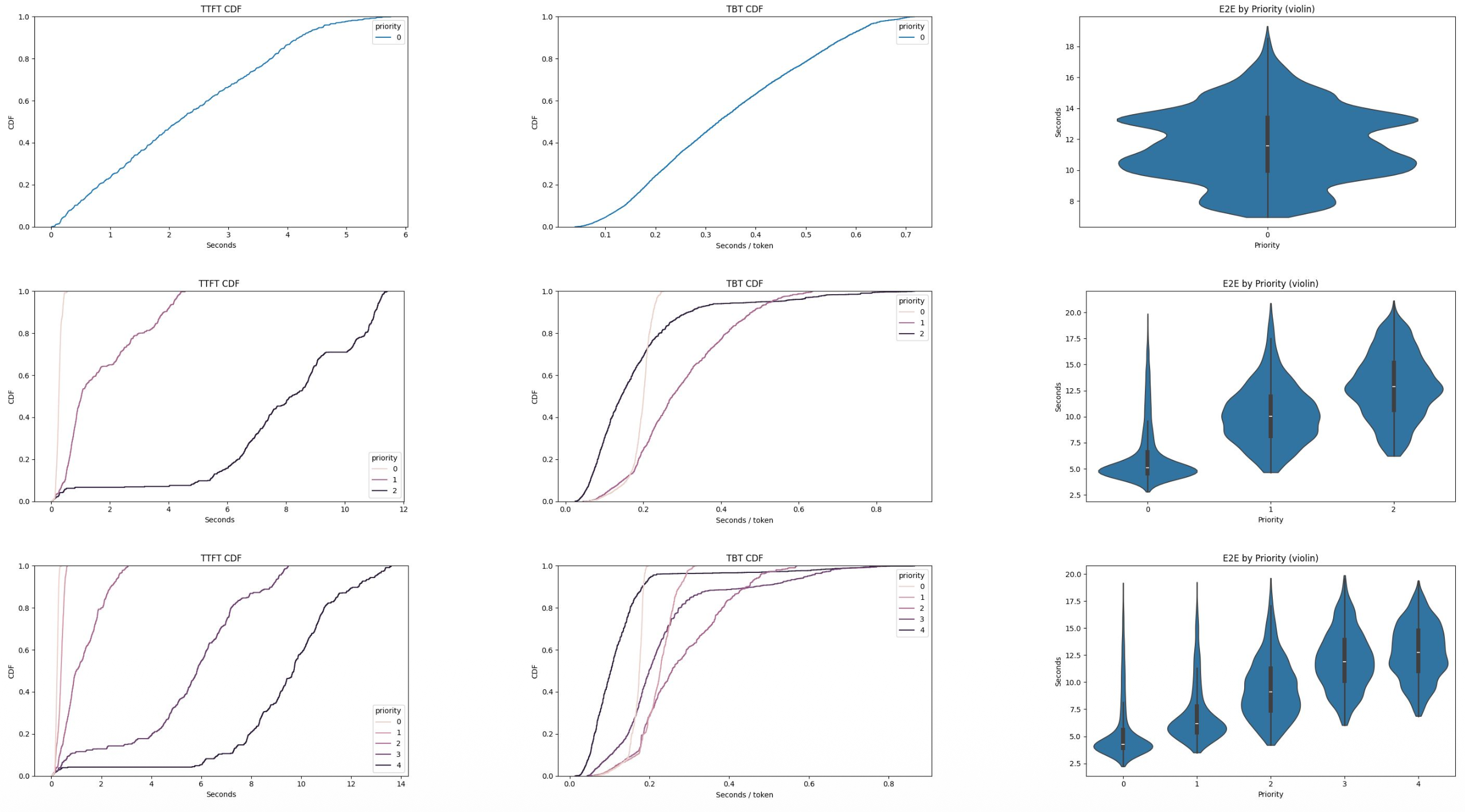}
\caption{Priority differentiation under \textbf{uniform} priority
    distribution. Rows: $K=1$, $K=3$, $K=5$ priority tiers. Columns: TTFT
    CDF, TBT CDF, end-to-end latency violin plot by priority.}
\label{fig:uniform}
\end{figure*}

\noindent\textbf{Uniform distribution.} Figure~\ref{fig:uniform} shows TTFT
CDF, TBT CDF, and end-to-end latency violin plots for $K \in \{1, 3, 5\}$
under the uniform distribution. With $K=1$ (no differentiation), all requests
experience identical latency with median end-to-end around 10--12 seconds.
With $K=3$, tier~0 achieves a median TTFT of approximately 0.3\,s versus over
3\,s for tier~2, demonstrating effective isolation. With $K=5$,
differentiation continues as tier~0 remains fast while tier~4 experiences the
longest tail. The gap between adjacent mid-priority tiers narrows, however,
and the overall system P99 increases slightly.

\noindent\textbf{Gaussian distribution.} The Gaussian distribution
concentrates requests at mid-priority tiers (Figure~\ref{fig:gaussian} in
Appendix~\ref{app:grids}). Tier~0 has few requests, so its headroom is rarely
fully consumed, resulting in very aggressive TTFT improvements (median TTFT
$<0.5$\,s for $K=5$, tier~0). The headroom budget is therefore less
efficiently utilized overall, as it remains idle for most of the experiment.
The tail behavior of lower tiers worsens significantly under $K=5$, consistent
with increasing headroom consumption.

\noindent\textbf{Enterprise distribution.} The enterprise distribution
(Figure~\ref{fig:enterprise} in Appendix~\ref{app:grids}) shows the most
practically relevant results. With 10\% critical traffic, tier~0 headroom is
consistently utilized, yielding stable TTFT improvements. The large fraction of
mid-priority traffic (tiers 1--2) benefits from isolation from background
traffic at tier $K{-}1$. This distribution consistently produces the largest
absolute speedup numbers (Section~\ref{sec:eval:speedup}).

\subsection{Scheduler Comparison Across Priority Levels}
\label{sec:eval:schedulers}
\begin{figure*}[t]
\centering
\setlength{\tabcolsep}{2pt}
\renewcommand{\arraystretch}{1.0}

\begin{tabular}{cccc}
{\footnotesize \textbf{$K=1$}} & {\footnotesize \textbf{$K=2$}} & {\footnotesize \textbf{$K=3$}} & {\footnotesize \textbf{$K=4$}} \\
\includegraphics[width=0.242\textwidth,trim=8 8 8 8,clip]{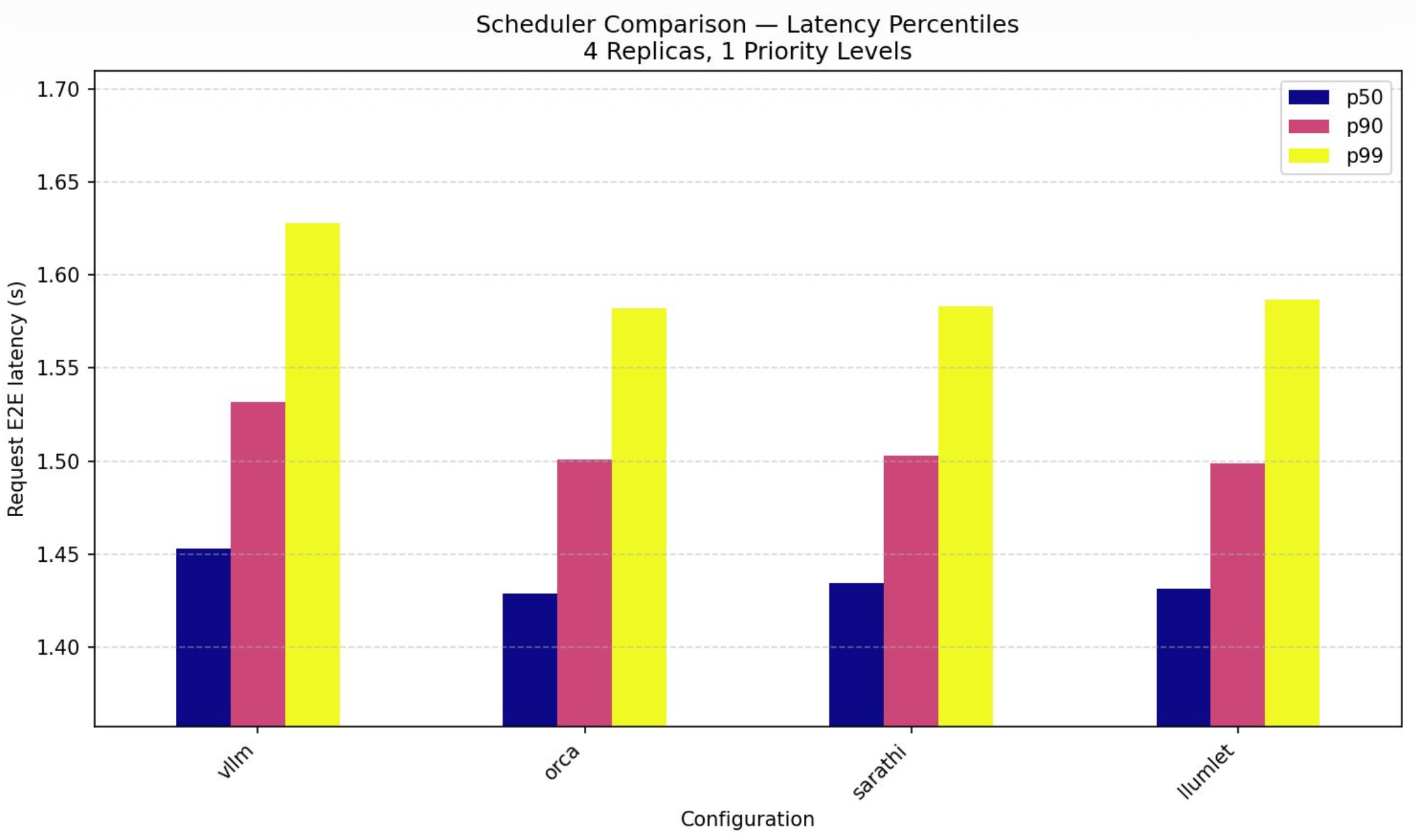} &
\includegraphics[width=0.242\textwidth,trim=8 8 8 8,clip]{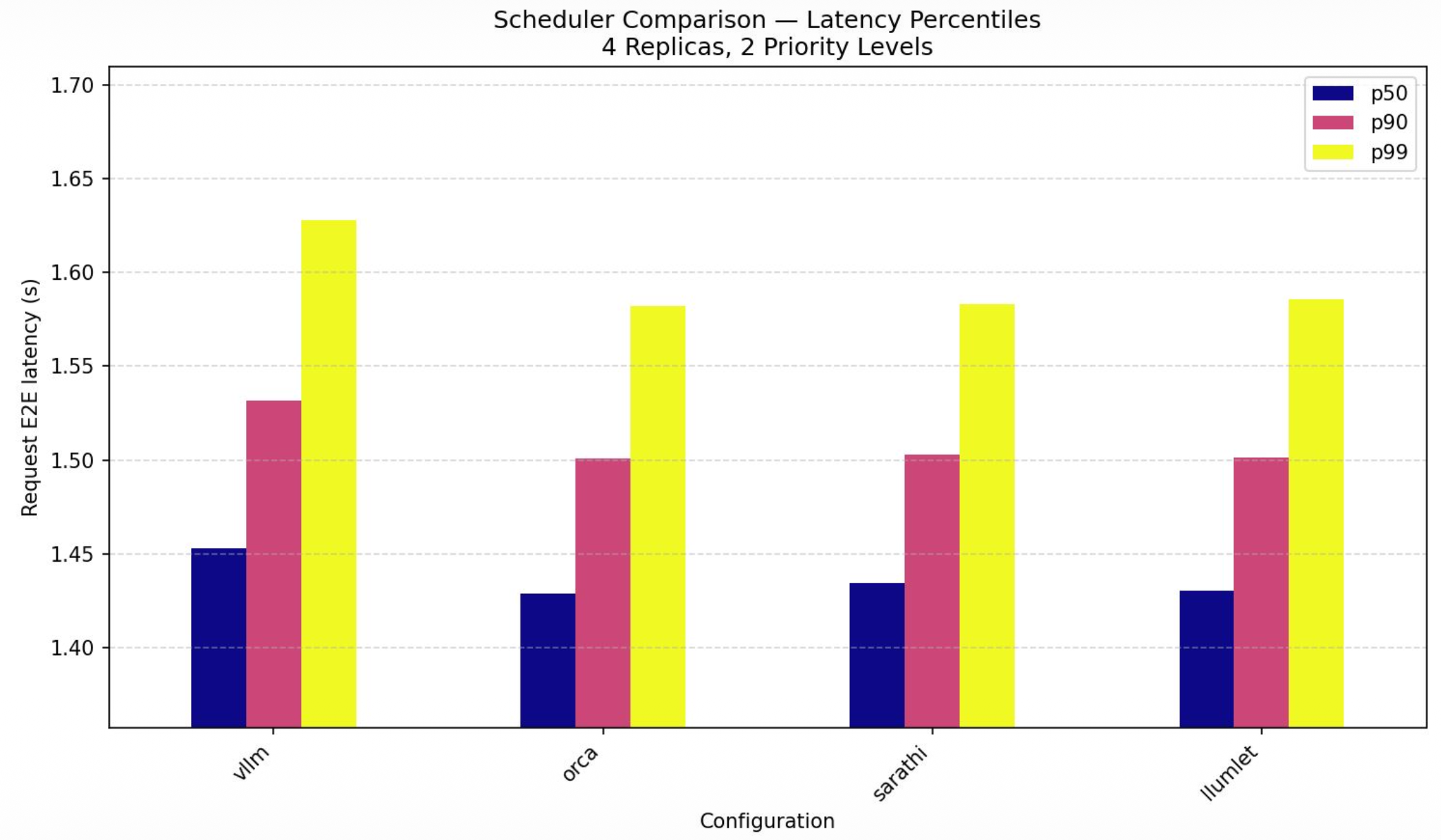} &
\includegraphics[width=0.242\textwidth,trim=8 8 8 4,clip]{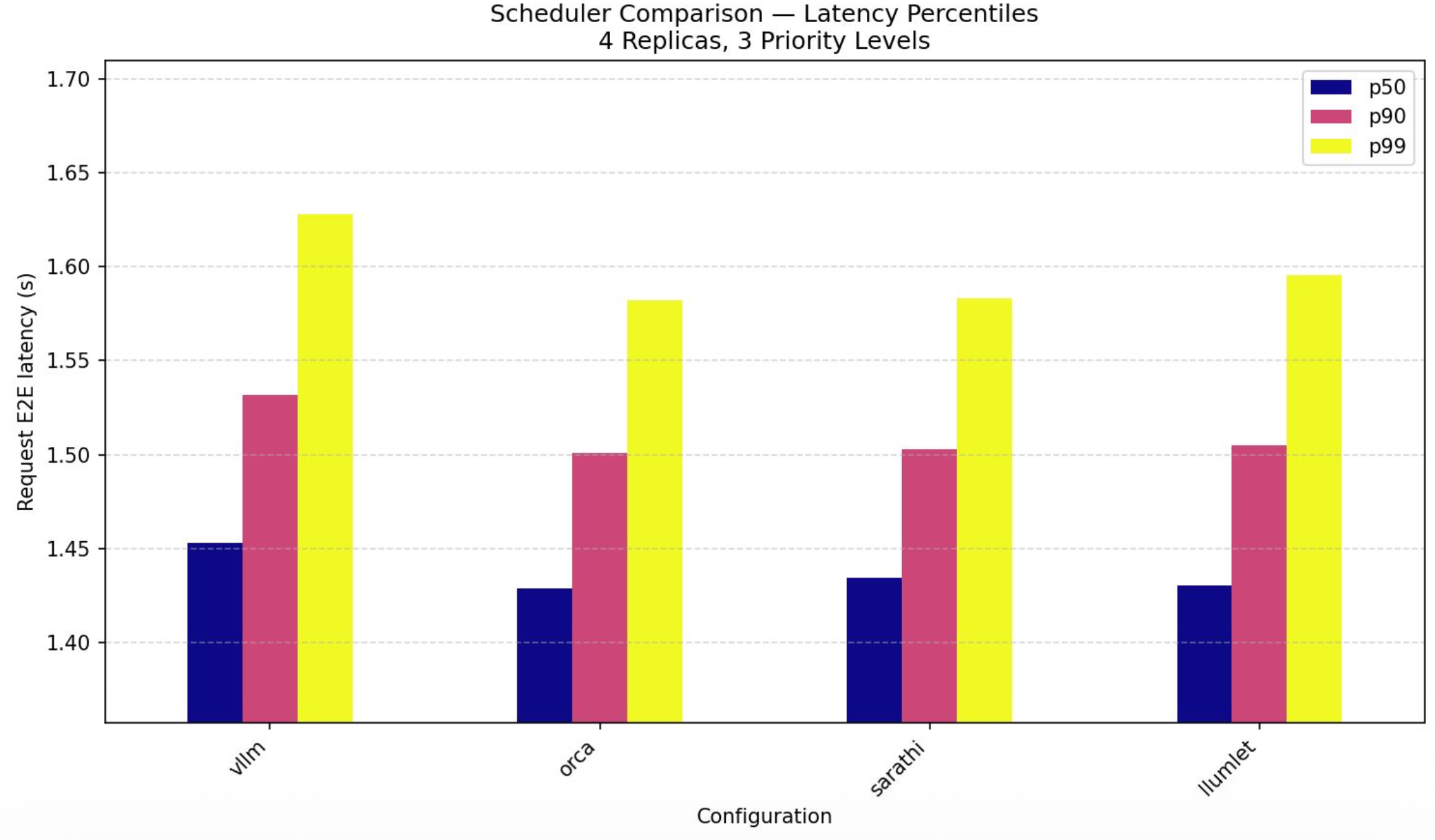} &
\includegraphics[width=0.242\textwidth,trim=8 8 8 8,clip]{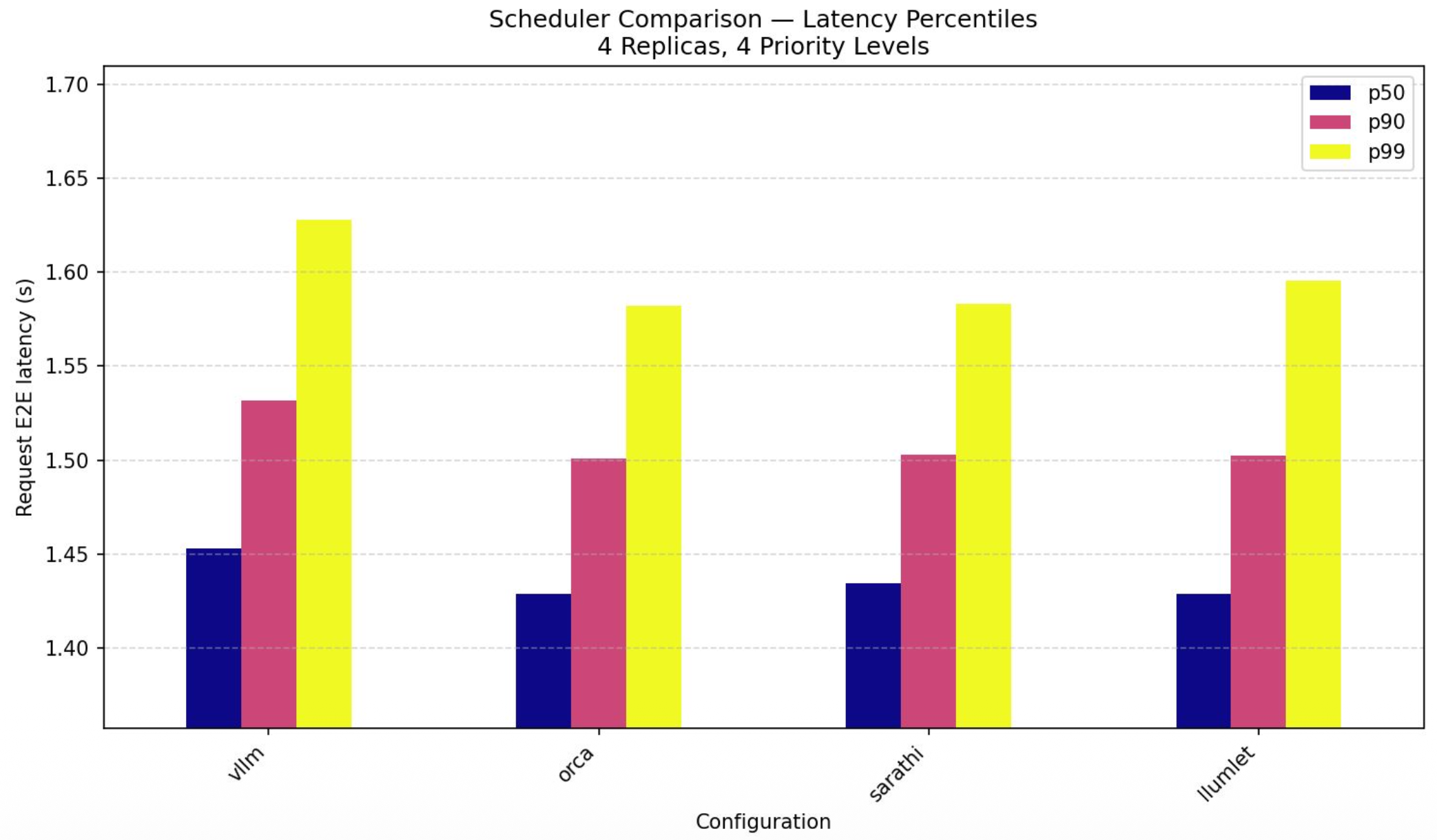} \\[3pt]

{\footnotesize \textbf{$K=5$}} & {\footnotesize \textbf{$K=6$}} & {\footnotesize \textbf{$K=7$}} & {\footnotesize \textbf{$K=8$}} \\
\includegraphics[width=0.242\textwidth,trim=8 8 8 4,clip]{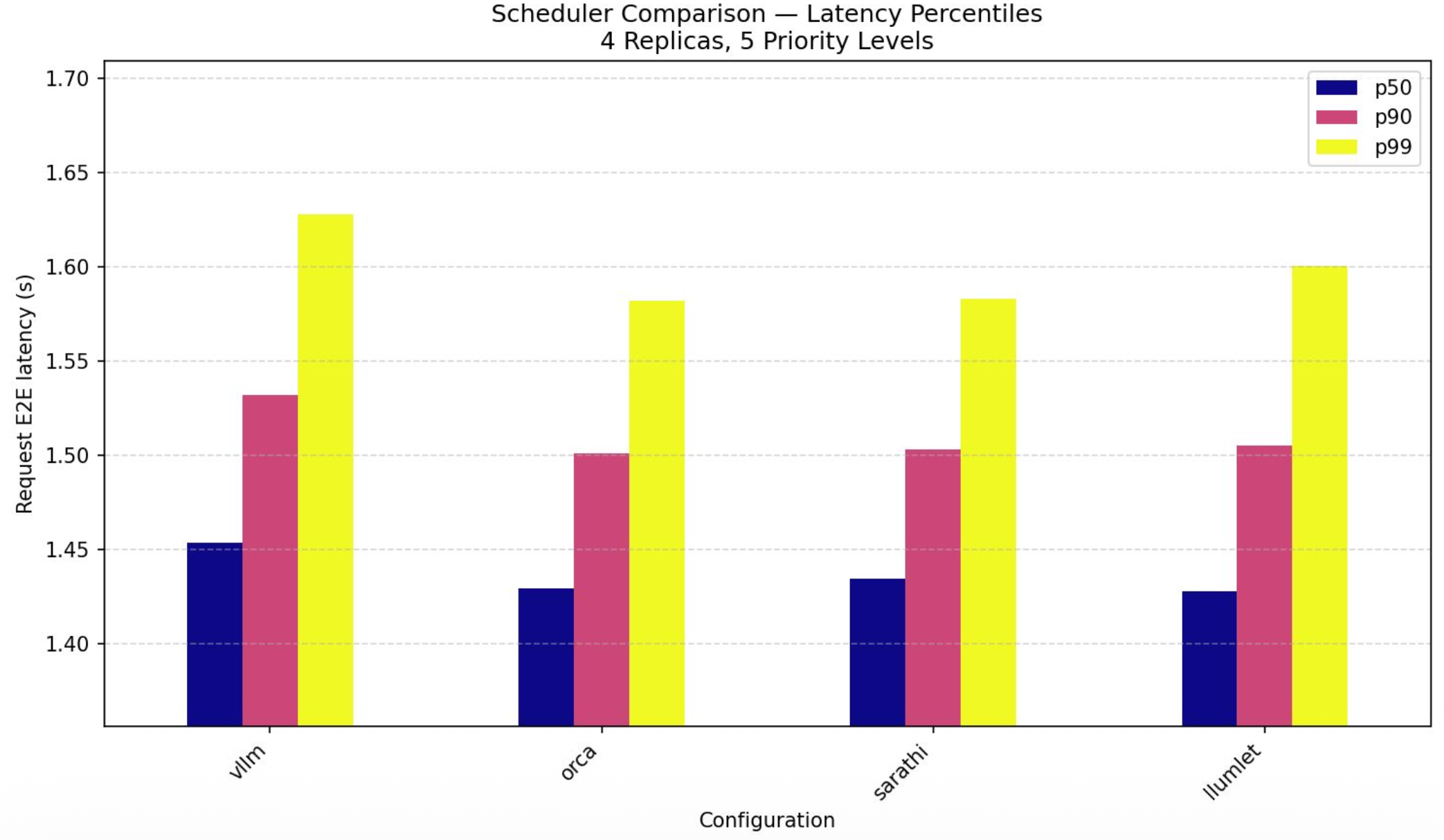} &
\includegraphics[width=0.242\textwidth,trim=8 8 8 4,clip]{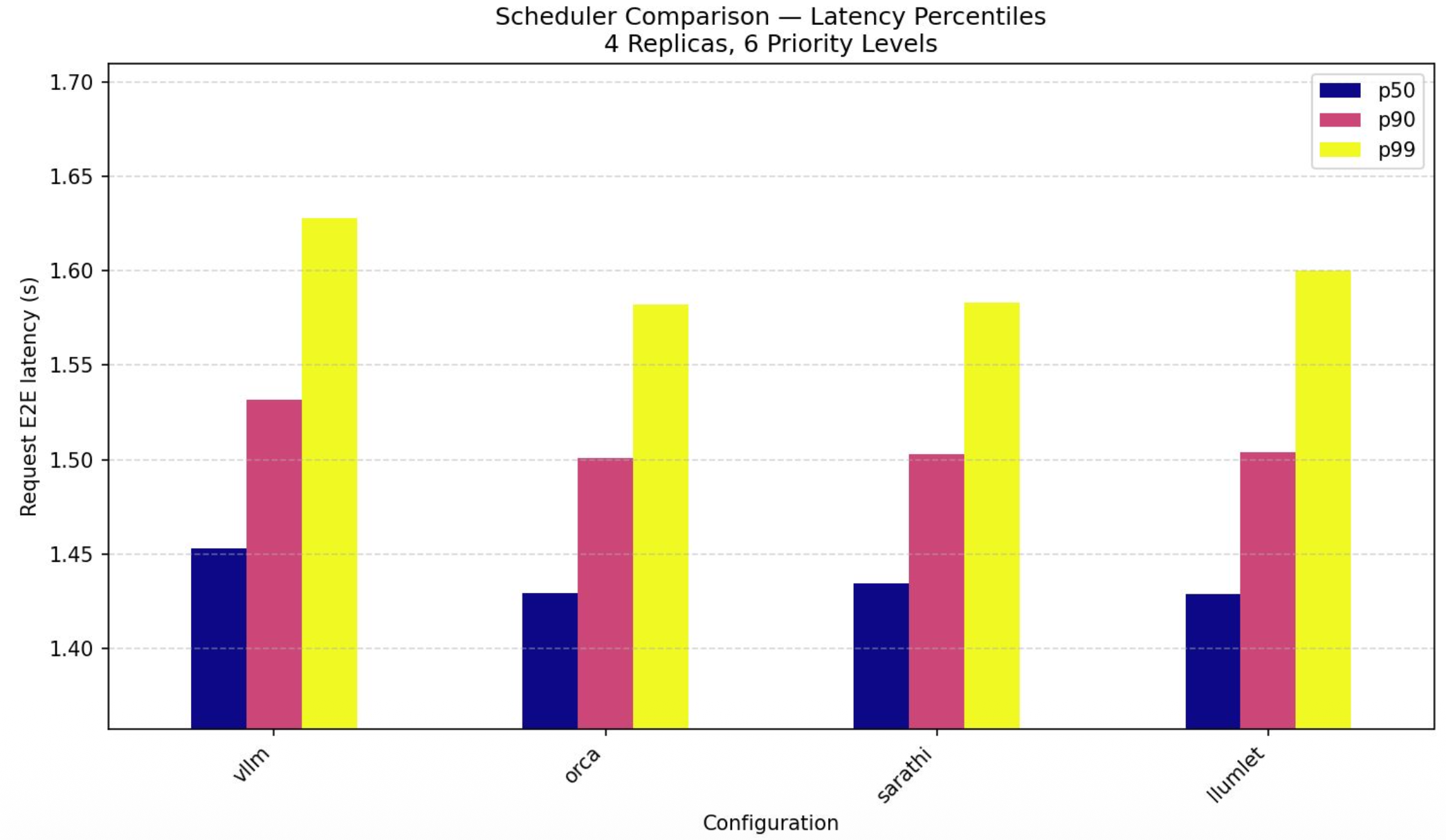} &
\includegraphics[width=0.242\textwidth,trim=8 8 8 8,clip]{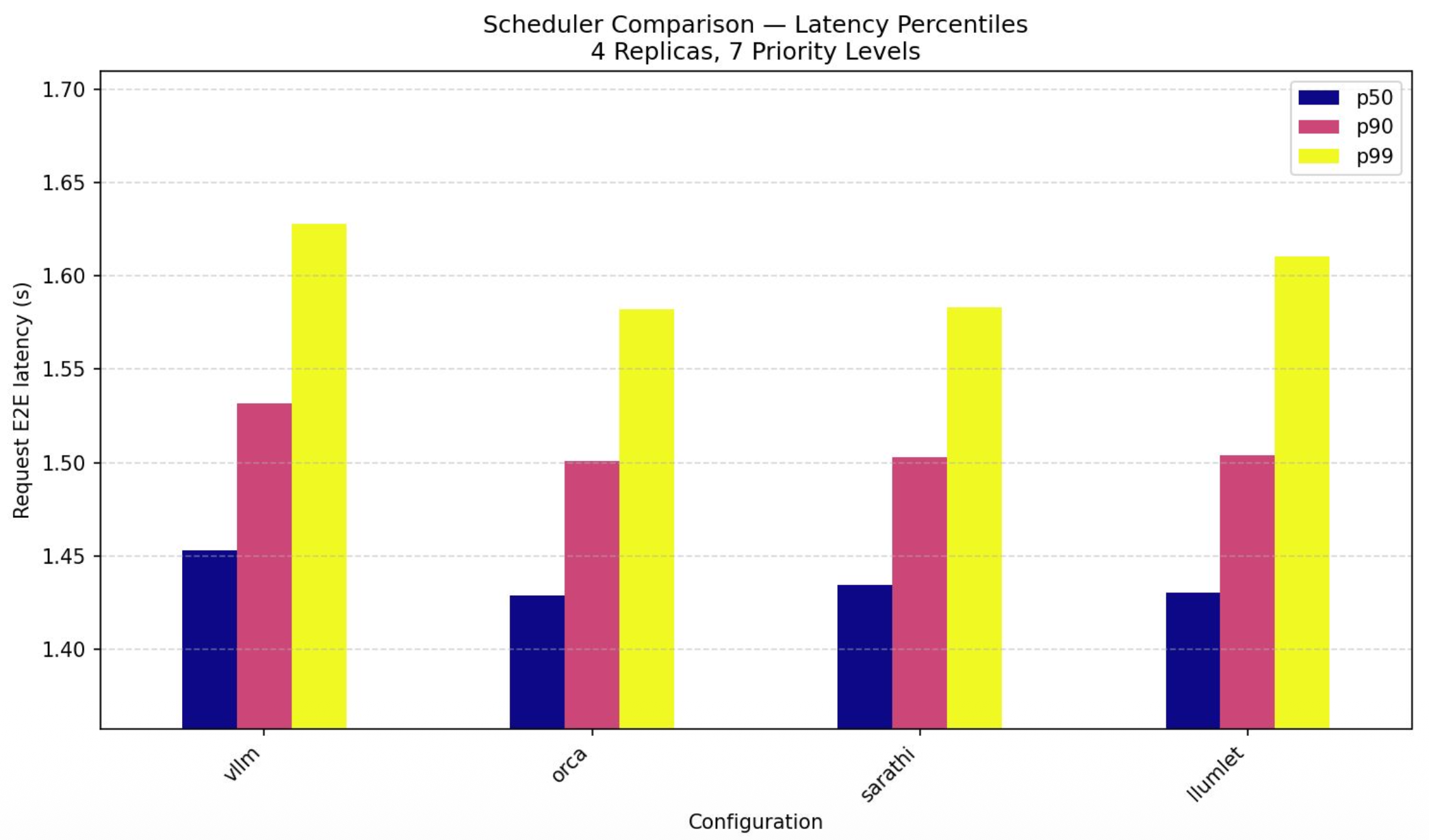} &
\includegraphics[width=0.242\textwidth,trim=8 8 8 4,clip]{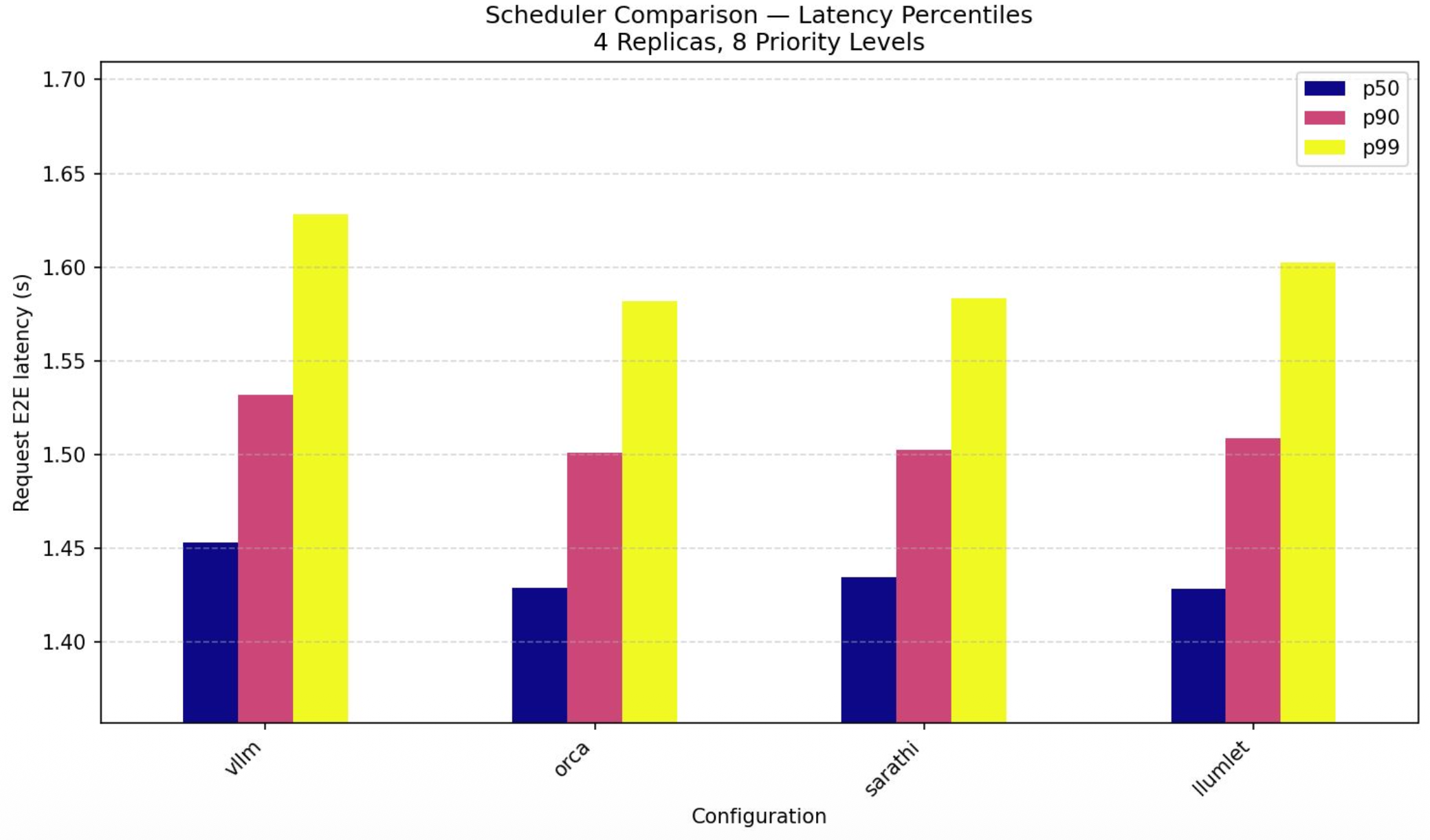} \\[3pt]

\multicolumn{4}{c}{
\begin{tabular}{cc}
{\footnotesize \textbf{$K=9$}} & {\footnotesize \textbf{$K=10$}} \\
\includegraphics[width=0.242\textwidth,trim=8 8 8 8,clip]{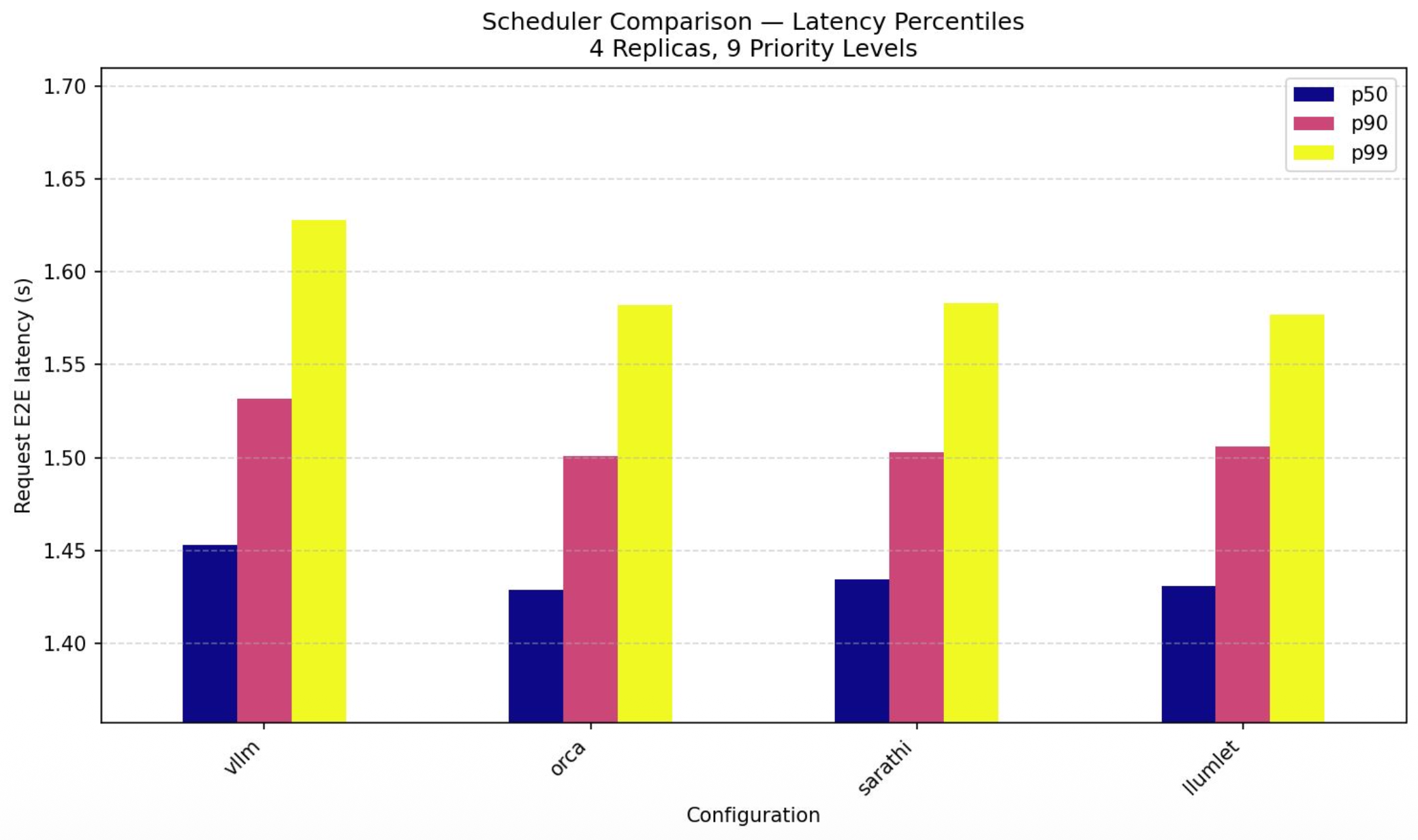} &
\includegraphics[width=0.242\textwidth,trim=8 8 8 8,clip]{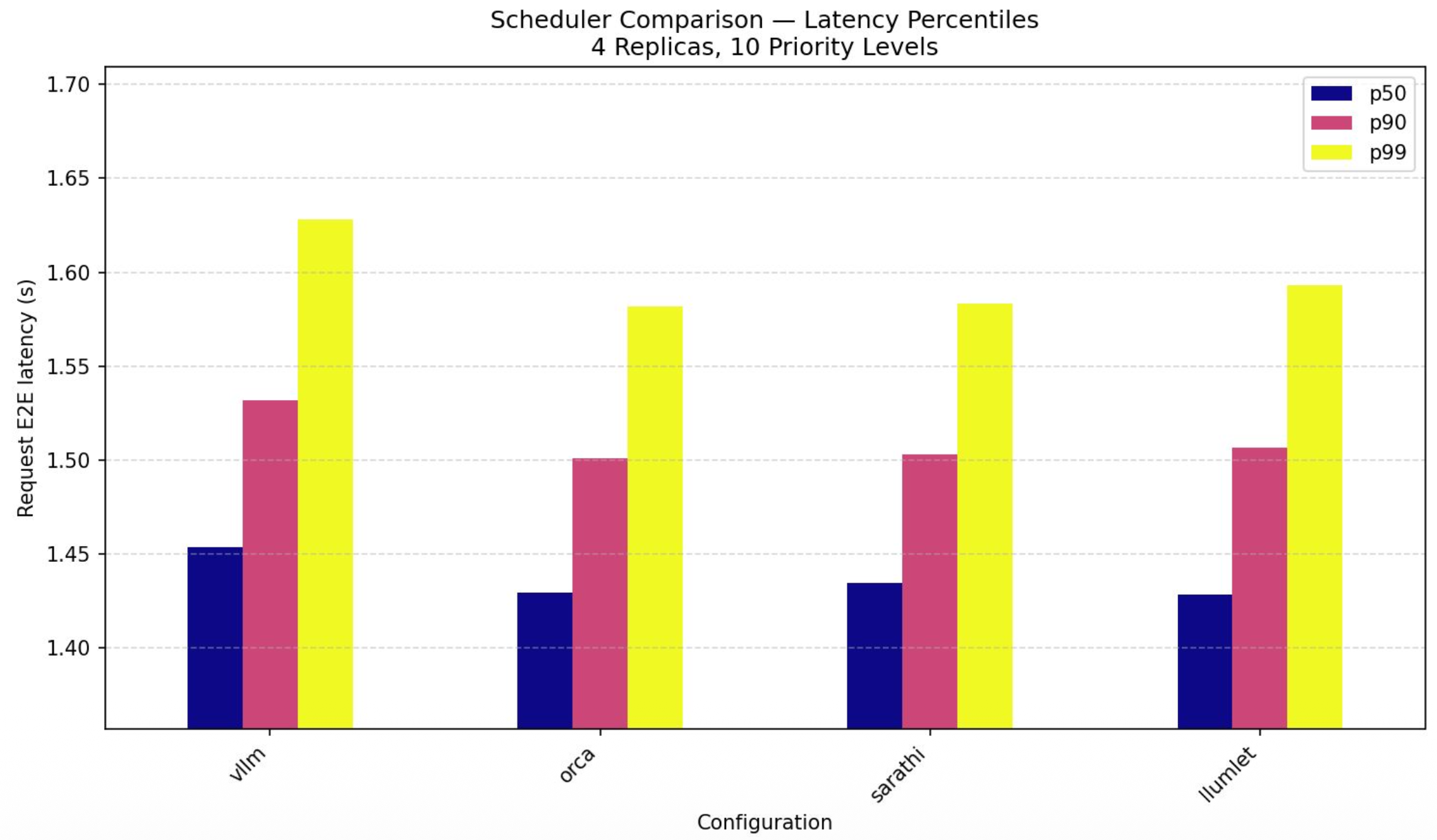}
\end{tabular}
}
\end{tabular}

\caption{End-to-end latency (P50/P90/P99) for all schedulers as a function of
priority level count $K$ (4 replicas, 1{,}000 requests, QPS $=10$). Across
$K=1$--$10$, aggregate latency remains broadly stable, indicating that increasing priority granularity does not introduce large overhead at this load.}
\label{fig:scheduler_comparison}
\end{figure*}

Figure~\ref{fig:scheduler_comparison} shows end-to-end latency percentiles
(P50, P90, P99) for all four replica schedulers (vLLM, Orca, Sarathi, Llumlet)
as priority levels increase from 1 to 10, with 4 replicas and 1{,}000
requests at QPS~$=10$. This benchmark is deliberately run at moderate load
to isolate cross-scheduler overhead; it is \emph{not} designed to demonstrate
per-tier latency differentiation, which is shown separately in
Section~\ref{sec:eval:differentiation}.

At $K=1$ (no priority differentiation), all schedulers perform similarly:
vLLM shows slightly higher P99 (1.63\,s) compared to Orca, Sarathi, and
Llumlet (1.58--1.59\,s). This confirms that baseline hardware and request
distributions are equivalent across configurations.

\begin{figure*}[t]
\centering
\includegraphics[width=\textwidth]{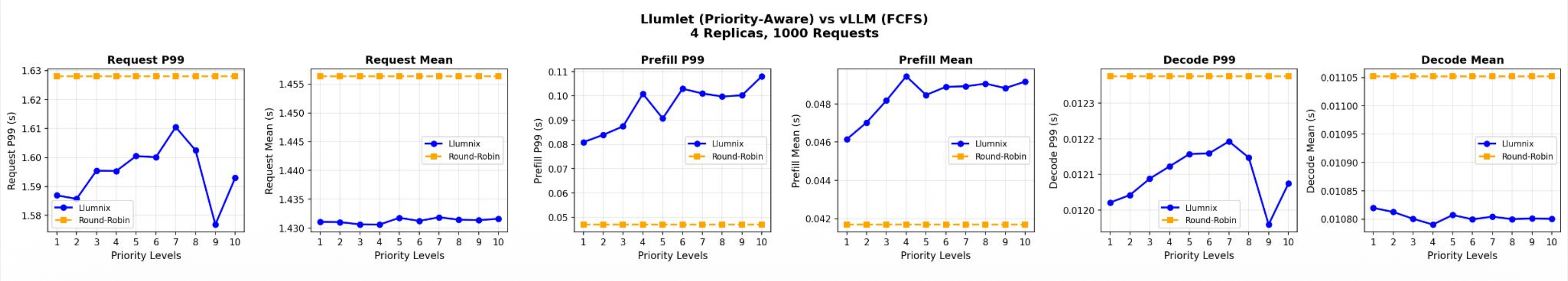}
\caption{Llumlet vs.\ vLLM Round-Robin across priority levels $K=1$--$10$,
    broken down by request end-to-end, prefill, and decode latency (P99 and
    mean).}
\label{fig:prefill_decode}
\end{figure*}

As $K$ increases, Llumlet's P99 rises modestly to $\sim$1.61\,s at $K=7$
before recovering, while vLLM, Orca, and Sarathi remain flat. This reflects
the cost of priority headroom: more active tiers increase total virtual
usage, reducing freeness and making the replica appear ``busier.'' Notably,
this overhead is concentrated in prefill latency
(Figure~\ref{fig:prefill_decode}), while decode latency remains stable and
slightly better than baselines throughout the sweep.

Importantly, the aggregate P50/P90/P99 across all tiers masks the per-tier
story. Under QPS~$=10$, the system operates below saturation and headroom
budgets rarely bind — all requests are served quickly regardless of tier,
so there is little headroom-driven differentiation to observe in the aggregate.
A heavily loaded workload would show high-priority tiers achieving lower
latency at the expense of low-priority tiers, widening the aggregate P99.
The value of Figure~\ref{fig:scheduler_comparison} is therefore to confirm
\emph{baseline parity}: Llumlet does not degrade aggregate performance
relative to simpler schedulers, a necessary condition for production
deployment.

\noindent\textbf{Key takeaway}: Llumlet supports up to 10 priority levels
without latency collapse and achieves aggregate baseline parity under
moderate load. Prioritization overhead is modest and primarily manifests in
prefill scheduling. Decode-phase performance is unaffected or slightly
improved, consistent with Llumnix's finding that migration mechanisms
disproportionately benefit long-running decode operations. The flat aggregate
trend across schedulers is expected at sub-saturation QPS and should not be
interpreted as null evidence for priority effectiveness; per-tier
differentiation is demonstrated under the controlled sweeps of
Section~\ref{sec:eval:differentiation}.

\subsection{Llumnix+Llumlet vs.\ INFaaS+vLLM}
\label{sec:eval:speedup}
Table~\ref{tab:uniform} (uniform distribution) and
Tables~\ref{tab:gaussian}--\ref{tab:enterprise} (Gaussian and enterprise
distributions; see Appendix~\ref{app:tables}) show speedup of our system
(Llumnix global + Llumlet local) over the INFaaS baseline (INFaaS global +
vLLM local) for $K \in \{3, 4, 5\}$ under three distributions at 10K and
15K requests.

% ---- Table 1 ----
\begin{table*}[t]
\centering
\caption{Speedup over INFaaS+vLLM --- \textbf{Uniform} distribution.
    Bold $K$ marks the best-performing configuration.}
\label{tab:uniform}
\small
\begin{tabular}{l r r r r r c}
\toprule
\textbf{Scale} & \textbf{Prefill P99} & \textbf{Prefill Mean}
& \textbf{E2E P99} & \textbf{E2E Mean} & \textbf{Cost/Lat.} & \textbf{K} \\
\midrule
\multirow{3}{*}{10K requests}
& 4.79$\times$ & 8.23$\times$ & 2.87$\times$ & 2.80$\times$ & 65\% & 3 \\
& 4.87$\times$ & 8.23$\times$ & 3.13$\times$ & 2.88$\times$ & 68\% & \textbf{4} \\
& 4.16$\times$ & 8.11$\times$ & 3.04$\times$ & 2.92$\times$ & 67\% & 5 \\
\midrule
\multirow{3}{*}{15K requests}
& 2.98$\times$ & 5.00$\times$ & 1.87$\times$ & 1.97$\times$ & 46\% & 3 \\
& 3.16$\times$ & 5.16$\times$ & 2.12$\times$ & 2.08$\times$ & 53\% & 4 \\
& 2.72$\times$ & 5.07$\times$ & 2.04$\times$ & 2.12$\times$ & 51\% & 5 \\
\midrule
    Llumnix~\cite{sun2024llumnix}
& $\leq$5.5$\times$ & $\leq$2.2$\times$
& \makecell[r]{$\leq$2.9$\times$ (real)\\$\leq$1.6$\times$ (synth.)}
& \makecell[r]{$\leq$2.0$\times$ (real)\\$\leq$1.6$\times$ (synth.)}
& 16--36\% & 2 \\
\bottomrule
\end{tabular}
\end{table*}

\noindent\textbf{Four priority tiers is consistently optimal.} At $K=4$, all
three distributions achieve peak E2E P99 speedup, peak or near-peak E2E mean
speedup, and peak cost-per-latency improvement. This sweet spot arises because
four tiers provides enough granularity to separate critical, high, standard,
and background traffic without fragmenting queues so finely that
load-balancing heuristics lose effectiveness. In addition, headroom budgets at
$K=4$ are still large enough to provide meaningful isolation for tiers~0
and~1, while the lower tiers remain efficiently batched.

\noindent\textbf{Prefill mean speedup exceeds the original paper.} Across all
conditions, our prefill mean speedup (5.0--8.3$\times$) substantially exceeds
Llumnix's reported $\leq$2.2$\times$. We attribute this to Llumlet's
migration-driven load balancing, which continuously redistributes prefill work
to underloaded replicas and prevents the ``convoy effect'' where bursty
prefill requests pile up on a single overloaded instance. INFaaS's reactive
cost model responds more slowly to transient prefill bursts.

\noindent\textbf{Prefill P99 speedup is lower at high load.} At 15K requests,
prefill P99 speedup drops to 2.0--3.2$\times$ compared to the paper's
reported $\leq$5.5$\times$. Near saturation, all replicas approach capacity
limits, reducing the freeness differential that drives migration. Prefill P99
is inherently more sensitive to bursty memory spikes than the mean, and
KV-cache pressure at the tail cannot be fully relieved when no destination has
sufficient slack.

\noindent\textbf{Cost-per-latency improvements significantly exceed the
original paper.} Our improvements of 46--68\% (10K) and 24--53\% (15K)
compare favorably to the paper's 16--36\%. This is attributable to the
migration system's ability to consolidate low-priority requests and free
capacity for high-priority workloads, resulting in less wasted KV-cache
memory and fewer preemption penalties.

% ===========================================================
\section{Discussion}
\label{sec:discussion}
% ===========================================================
\subsection{The Priority-Granularity Tradeoff}
\label{sec:disc:tradeoff}
Our results support a general principle we term the \emph{priority-granularity
tradeoff}: as the number of priority tiers $K$ increases, two opposing effects
compete. Finer granularity enables more precise SLA differentiation, but also
increases total headroom consumption, since each tier with any active requests
contributes its full budget $H_p$ to virtual usage. This in turn reduces
effective batching capacity and increases average latency for all tiers.

The empirically observed optimum at $K=4$ is robust across workload
distributions. For future deployments, we recommend starting at 3--4 priority
tiers and increasing only if SLO differentiation requirements cannot otherwise
be met, rather than deploying maximum granularity by default.

\subsection{Interaction Between Load and Priority Effectiveness}
\label{sec:disc:load}
Benefits of priority-aware scheduling are most pronounced at moderate load
(10K requests), where freeness variance across replicas is higher and
migration opportunities are abundant. Near saturation (15K requests), the
average freeness approaches zero and the system transitions from a regime
where migration can rebalance load to one where all replicas are equally
overloaded. This suggests that priority scheduling should be paired with
proactive auto-scaling: detecting saturation early and scaling out before the
migration mechanism loses effectiveness.

A related subtlety concerns the measurement of priority effectiveness itself.
Aggregate latency percentiles (pooled across all tiers) are an incomplete
metric: even when Llumlet is working correctly, aggregate P99 may be similar
to or slightly higher than baselines because high-priority gains are offset by
low-priority degradation. The correct metric is \emph{per-tier} latency, which
would show tier-0 improving relative to baselines while tier $K{-}1$ worsens
--- exactly the intended SLA tradeoff. The per-tier differentiation plots
(Figures~\ref{fig:uniform}--\ref{fig:enterprise}) capture this, while the
aggregate scheduler comparison (Figure~\ref{fig:scheduler_comparison})
measures a different property: aggregate overhead neutrality.

\subsection{Distribution-Specific Behavior}
\label{sec:disc:distributions}
The enterprise distribution consistently yields the best cost-efficiency
results due to its skewed priority assignment. The concentration of requests
at mid-priority tiers creates well-populated queues that Llumnix's batch
normalization can exploit effectively, while the small fraction of tier-0
traffic means that critical requests rarely compete with each other for the
headroom budget. This mirrors real-world deployments where ``VIP'' traffic is
a small fraction of total volume.

The Gaussian distribution shows the highest sensitivity to tier count, with
speedups varying significantly between $K=3$ and $K=5$. Schedulers deployed
on Gaussian workloads (e.g., consumer APIs where most users are ``standard'')
may benefit from workload-adaptive tier reconfiguration.

\subsection{Comparison to the Original Llumnix Paper}
\label{sec:disc:comparison}
Our simulation-based results cannot be directly compared to Llumnix's
hardware-measured results~\cite{sun2024llumnix} for three reasons. First, we
implemented our own versions of Llumnix and INFaaS adapted to Vidur's
architecture, which may differ in subtle ways from the original
implementations. Second, Vidur simulates GPU computation without capturing
all hardware effects such as PCIe bandwidth for KV migration and CUDA kernel
launch overhead. Third, we did not have access to the exact workload traces
used in the original paper. Nonetheless, our results are broadly consistent
with the paper's reported ranges, and in some metrics (prefill mean,
cost-per-latency) exceed them, lending confidence that our implementation
captures the essential dynamics of Llumnix's scheduling design.

\subsection{Limitations and Future Work}
\label{sec:disc:future}
\noindent\textbf{Adaptive headroom allocation.} Our exponential decay schedule
is static. A natural extension is adaptive headroom that continuously monitors
the observed latency differential between tiers, adjusting $H_p$ upward when
SLO violations are detected and downward when headroom is consistently
unused~\cite{tang2025scorpio,chen2025slosserve}.

\noindent\textbf{Multilevel feedback queues.} Tier assignment is fixed at
request arrival in our system. An MLFQ-style
extension~\cite{corbato1962mlfq} could dynamically promote requests that have
waited longer than a tier-specific time quota, preventing starvation of
low-priority requests under sustained high load.

\noindent\textbf{Vidur-Search integration.} Vidur's configuration search
capability~\cite{agrawal2024vidur} could automatically find optimal headroom
schedules and tier counts for a given workload distribution and SLO target,
replacing manual tuning with systematic exploration.

\noindent\textbf{Contention level and per-tier evaluation.} The scheduler
comparison (Figure~\ref{fig:scheduler_comparison}) runs at QPS~$=10$, placing
the system below saturation. At this load, headroom budgets rarely bind and
aggregate latency percentiles pooled across all tiers are similar for all
schedulers. Demonstrating the expected per-tier latency split --- high-priority
tiers improving at the expense of low-priority tiers --- would require both
higher utilization and reporting per-tier breakdowns separately rather than
pooled aggregates. This is a key direction for future evaluation and
represents the natural next step to validate the priority differentiation
results of Section~\ref{sec:eval:differentiation} at realistic production
load.

\noindent\textbf{Hardware validation.} Validating our simulation results on a
real GPU cluster (e.g., A100 or H100 instances) would quantify the gap
between simulated and hardware performance, particularly for migration costs,
which depend critically on PCIe and NVLink bandwidth.

% ===========================================================
\section{Conclusion}
\label{sec:conclusion}
% ===========================================================
We have presented a multi-tier extension of Llumnix's priority scheduling
model, implemented and evaluated within the Vidur LLM inference simulator.
By replacing Llumnix's binary high/normal priority classification with
$K$-tier scheduling backed by exponentially decayed per-tier headroom,
tier-aware dispatch ordering, and full migration support, we demonstrate that
meaningful SLO differentiation across four or more priority classes is
achievable without significant overall system throughput loss.

Our evaluation across three workload distributions, two request volume scales,
and priority levels ranging from 1 to 10 reveals a consistent optimum at
$K=4$. Four priority tiers achieves peak E2E P99 speedups of up to
$3.13{\times}$ and cost-per-latency improvements of up to 68\% over
INFaaS+vLLM, substantially outperforming both the original Llumnix paper's
reported cost savings and INFaaS's load-balancing approach. Beyond $K=5$,
benefits plateau and overhead from headroom fragmentation begins to dominate.

These results provide practical guidance for production LLM serving
deployments: a four-tier SLA model covering critical, high, standard, and
background traffic is both sufficient to capture the full spectrum of user
latency requirements and efficient enough to preserve the high-priority
optimizations that migration-capable schedulers provide.

% ===========================================================
\section*{Acknowledgments}
% ===========================================================
We thank the authors of the
Vidur and Llumnix open-source projects for making their frameworks publicly
available.

% ===========================================================
% Appendix  — switch to single column so headings, figures, and
%             tables render at full page width without hyphenation
% ===========================================================
\clearpage
\onecolumn
\appendix
% Make section headings read "Appendix A: Title" instead of just "A  Title"
\renewcommand{\thesection}{Appendix~\Alph{section}:}

% -------------------------------------------------------------------
\section{Priority Differentiation: Gaussian and Enterprise Grids}
\label{app:grids}
% -------------------------------------------------------------------

\begin{figure}[h!]
\centering
\includegraphics[width=\textwidth]{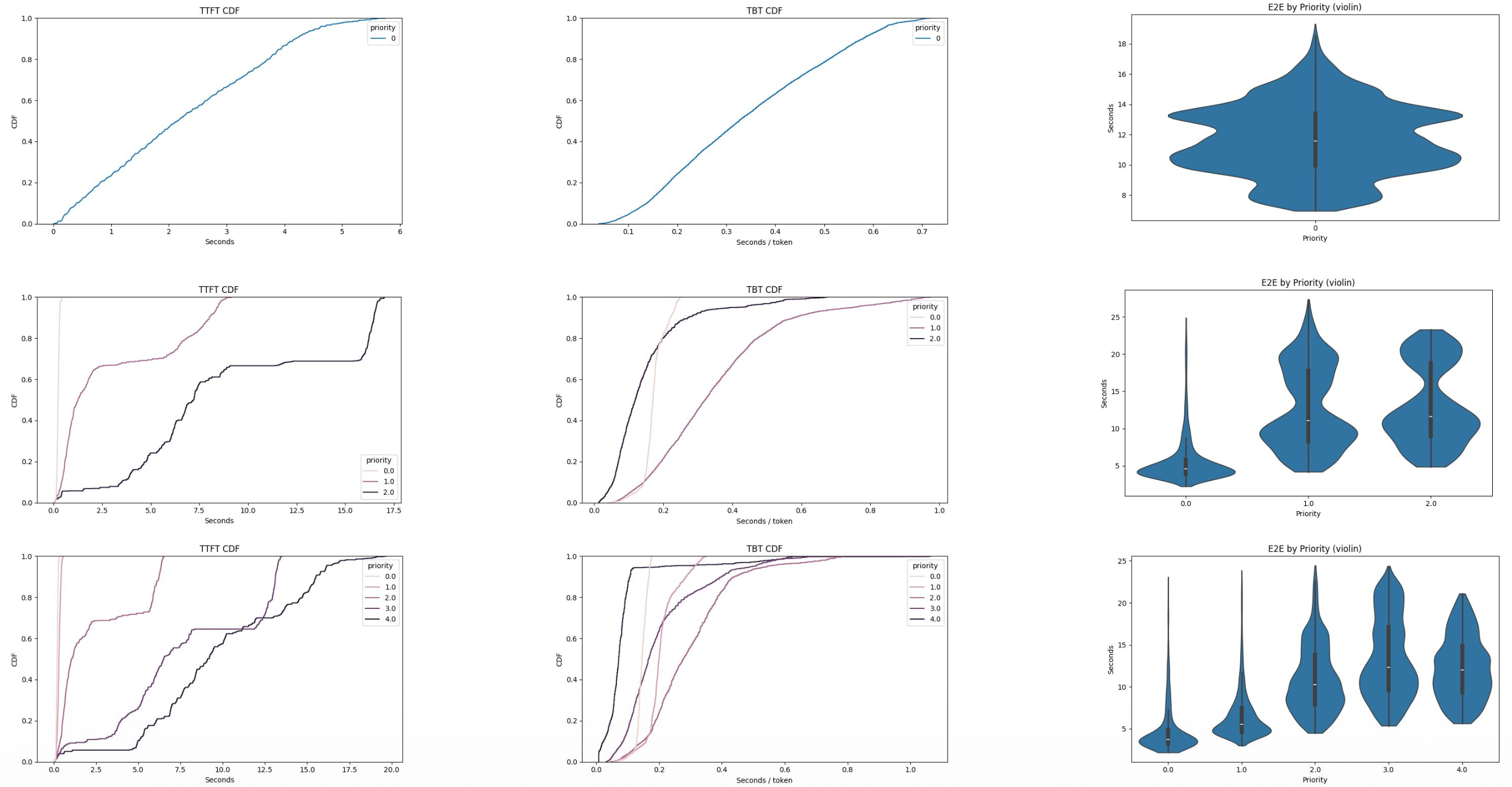}
\caption{Priority differentiation under \textbf{Gaussian} priority
    distribution. Rows: $K=1$, $K=3$, $K=5$. Columns: TTFT CDF, TBT CDF,
    end-to-end latency violin plot by priority.}
\label{fig:gaussian}
\end{figure}

\begin{figure}[h!]
\centering
\includegraphics[width=\textwidth]{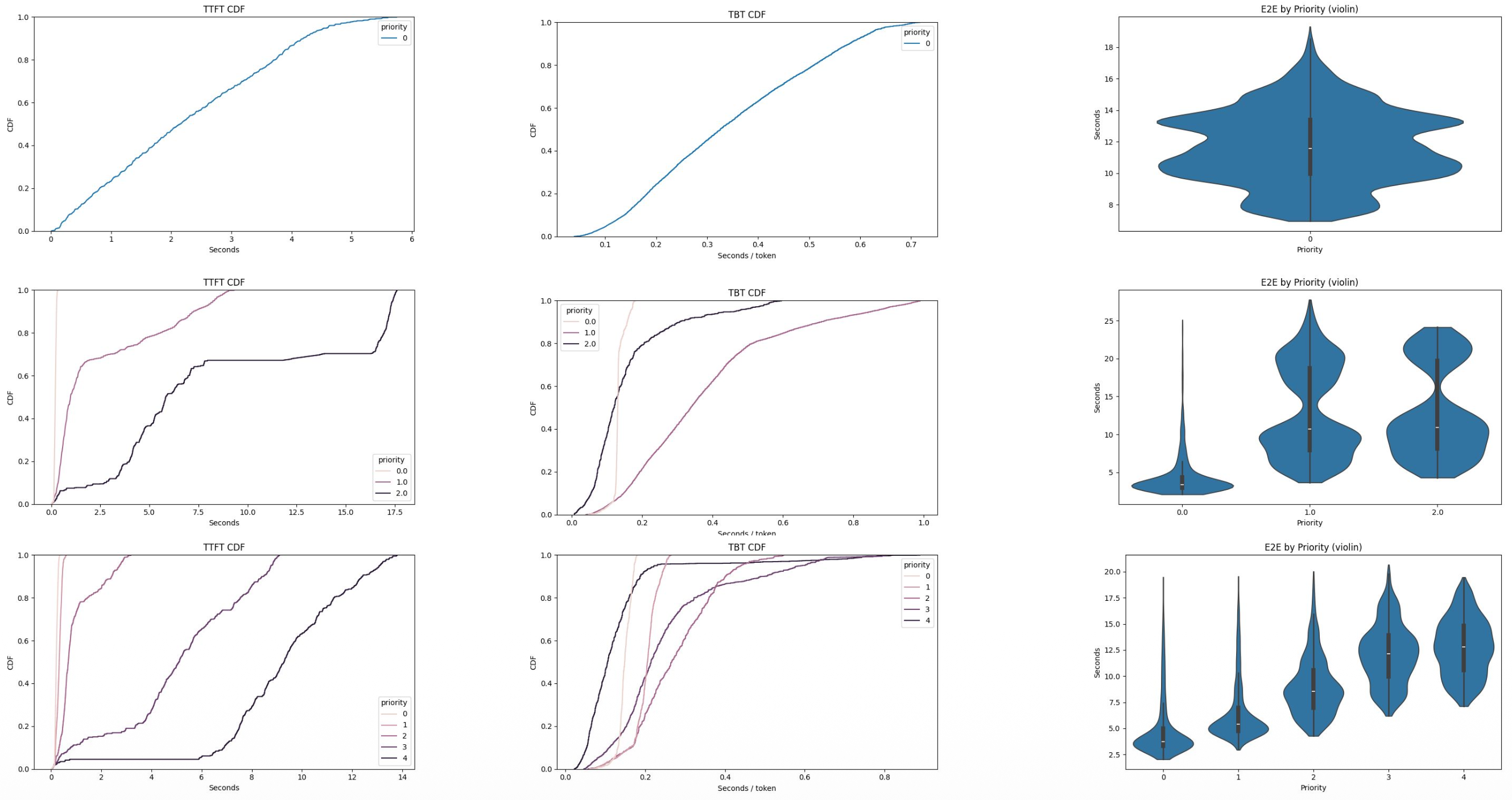}
\caption{Priority differentiation under \textbf{enterprise} priority
    distribution. Rows: $K=1$, $K=3$, $K=5$. Columns: TTFT CDF, TBT CDF,
    end-to-end latency violin plot by priority.}
\label{fig:enterprise}
\end{figure}

Figures~\ref{fig:gaussian} and~\ref{fig:enterprise} show the full TTFT CDF,
TBT CDF, and end-to-end latency violin plots for the Gaussian and enterprise
priority distributions, mirroring the format of Figure~\ref{fig:uniform} in
the main paper.

% -------------------------------------------------------------------
\FloatBarrier
\section{Speedup Tables: Gaussian and Enterprise Distributions}
\label{app:tables}
% -------------------------------------------------------------------

Tables~\ref{tab:gaussian} and~\ref{tab:enterprise} report speedup of
Llumnix+Llumlet over INFaaS+vLLM under the Gaussian and enterprise priority
distributions, mirroring the format of Table~\ref{tab:uniform} in the main
paper.  The $K=4$ optimum holds across all three distributions.

\begin{table}[h!]
\centering
\caption{Speedup over INFaaS+vLLM --- \textbf{Gaussian} distribution.
  Bold $K$ marks the best-performing configuration.}
\label{tab:gaussian}
\small
\begin{tabular}{l r r r r r c}
\toprule
\textbf{Scale} & \textbf{Prefill P99} & \textbf{Prefill Mean}
& \textbf{E2E P99} & \textbf{E2E Mean} & \textbf{Cost/Lat.} & \textbf{K} \\
\midrule
\multirow{3}{*}{10K requests}
& 3.24$\times$ & 7.47$\times$ & 2.26$\times$ & 2.45$\times$ & 56\% & 3 \\
& 4.25$\times$ & 8.33$\times$ & 3.07$\times$ & 2.79$\times$ & 67\% & \textbf{4} \\
& 3.49$\times$ & 7.51$\times$ & 2.43$\times$ & 2.74$\times$ & 59\% & 5 \\
\midrule
\multirow{3}{*}{15K requests}
& 2.00$\times$ & 4.68$\times$ & 1.49$\times$ & 1.71$\times$ & 33\% & 3 \\
& 2.70$\times$ & 5.24$\times$ & 2.02$\times$ & 1.96$\times$ & 51\% & 4 \\
& 2.25$\times$ & 4.88$\times$ & 1.97$\times$ & 1.67$\times$ & 41\% & 5 \\
\midrule
Llumnix~\cite{sun2024llumnix}
& $\leq$5.5$\times$ & $\leq$2.2$\times$
& \makecell[r]{$\leq$2.9$\times$ (real)\\$\leq$1.6$\times$ (synth.)}
& \makecell[r]{$\leq$2.0$\times$ (real)\\$\leq$1.6$\times$ (synth.)}
& 16--36\% & 2 \\
\bottomrule
\end{tabular}
\end{table}

\begin{table}[h!]
\centering
\caption{Speedup over INFaaS+vLLM --- \textbf{Enterprise} distribution.
  Bold $K$ marks the best-performing configuration.}
\label{tab:enterprise}
\small
\begin{tabular}{l r r r r r c}
\toprule
\textbf{Scale} & \textbf{Prefill P99} & \textbf{Prefill Mean}
& \textbf{E2E P99} & \textbf{E2E Mean} & \textbf{Cost/Lat.} & \textbf{K} \\
\midrule
\multirow{3}{*}{10K requests}
& 3.10$\times$ & 8.06$\times$ & 2.18$\times$ & 2.27$\times$ & 54\% & 3 \\
& 4.41$\times$ & 8.28$\times$ & 3.02$\times$ & 2.79$\times$ & 67\% & \textbf{4} \\
& 4.12$\times$ & 8.14$\times$ & 2.96$\times$ & 2.89$\times$ & 66\% & 5 \\
\midrule
\multirow{3}{*}{15K requests}
& 1.77$\times$ & 4.51$\times$ & 1.31$\times$ & 1.44$\times$ & 24\% & 3 \\
& 2.76$\times$ & 5.04$\times$ & 1.94$\times$ & 1.95$\times$ & 48\% & 4 \\
& 2.61$\times$ & 5.06$\times$ & 1.97$\times$ & 2.08$\times$ & 49\% & 5 \\
\midrule
Llumnix~\cite{sun2024llumnix}
& $\leq$5.5$\times$ & $\leq$2.2$\times$
& \makecell[r]{$\leq$2.9$\times$ (real)\\$\leq$1.6$\times$ (synth.)}
& \makecell[r]{$\leq$2.0$\times$ (real)\\$\leq$1.6$\times$ (synth.)}
& 16--36\% & 2 \\
\bottomrule
\end{tabular}
\end{table}

% ===========================================================
% Bibliography
% ===========================================================
\clearpage
\balance

\end{document}